%% file: main.tex
\documentclass[sigconf, screen]{acmart}

\setcopyright{rightsretained}
\acmDOI{10.1145/3650212.3680370}
\acmYear{2024}
\copyrightyear{2024}
\acmISBN{979-8-4007-0612-7/24/09}
\acmConference[ISSTA '24]{Proceedings of the 33rd ACM SIGSOFT International Symposium on Software Testing and Analysis}{September 16--20, 2024}{Vienna, Austria}
\acmBooktitle{Proceedings of the 33rd ACM SIGSOFT International Symposium on Software Testing and Analysis (ISSTA '24), September 16--20, 2024, Vienna, Austria}
\acmSubmissionID{issta24main-p1132-p}
\received{2024-04-12}
\received[accepted]{2024-07-03}
\usepackage{calc}
\usepackage[inline]{enumitem}
\usepackage{booktabs}
\usepackage{multirow}
\usepackage{venndiagram}
\usepackage{balance}

\usepackage{tikz}
\usetikzlibrary{shapes.geometric, shapes.symbols, arrows, calc}

\usepackage{graphicx}
\graphicspath{ {./images/} }
\usepackage[skip=1pt]{caption}
\usepackage[skip=1pt]{subcaption}

\usepackage{metalogo}
\usepackage{xspace}

\newcommand{\tex}{\TeX\xspace}
\newcommand{\latex}{\LaTeX\xspace}
\newcommand{\xetex}{\XeTeX\xspace}
\newcommand{\luatex}{\LuaTeX\xspace}
\newcommand{\pdftex}{pdf\TeX\xspace}
\newcommand{\pdflatex}{pdf\LaTeX\xspace}
\newcommand{\xelatex}{\XeLaTeX\xspace}
\newcommand{\lualatex}{\LuaLaTeX\xspace}
\newcommand{\texlive}{\TeX\xspace Live\xspace}

\usepackage{wasysym}
\newcommand{\yes}{\Circle}%
\newcommand{\no}{\CIRCLE}%

\begin{document}

% <ARTICLE SETUP>
\title{Inconsistencies in \tex-Produced Documents}

\author{Jovyn Tan}
\orcid{0009-0004-5086-9959}
\affiliation{%
  \institution{National University of Singapore}
  \city{Singapore}
  \country{Singapore}
}
\email{jovyn.tls@u.nus.edu}

\author{Manuel Rigger}
\orcid{0000-0001-8303-2099}
\affiliation{%
  \institution{National University of Singapore}
  \city{Singapore}
  \country{Singapore}
}
\email{rigger@nus.edu.sg}
% </ARTICLE SETUP>

%
\begin{abstract}
  \input{sections/abstract.tex}
\end{abstract}




\maketitle

% HACK: vertically shorten tables
% \renewcommand{\arraystretch}{0.9}

\input{sections/introduction.tex}
\input{sections/background.tex}

\input{sections/motivating-study.tex}

\input{sections/methodology.tex}

\input{sections/results-1.tex}

\input{sections/results-2.tex}

\input{sections/results-3.tex}

\input{sections/threats-to-validity.tex}

\input{sections/discussion.tex}

\input{sections/related-work.tex}

\input{sections/conclusion.tex}

\input{sections/data-availability.tex}

\bibliographystyle{ACM-Reference-Format}
\balance{}
\bibliography{main}

\end{document}

\begin{center}
  \Huge \textbf{Supplementary Materials}

  \huge for \textit{``Inconsistencies in \tex-Produced Documents''}
\end{center}

\appendix

\title{Supplementary Materials}

\tableofcontents

% -----------------------------------------------------------------------------

\section{Motivating Study}

\subsection{Sampled Documents}

\begin{table}[H]
  \caption{Documents sampled in the motivating study}
  \begin{tabular}{lll}\toprule 
    % & \multicolumn{2}{c}{Comparing \xetex with:} \\ \cmidrule(lr){2-3}
    \textbf{arXiv ID} & \textbf{Document class} & \textbf{Taxonomy} \\\midrule
    2306.00001  & \texttt{IEEEtran}       & cs.CV \\
    2306.00002  & \texttt{elsarticle}     & physics.soc-ph \\
    2306.00004  & \texttt{llncs}          & cs.SE \\
    2306.00022  & \texttt{mnras}          & astro-ph.EP \\
    2306.00036  & \texttt{article}        & cs.AI \\
    2306.00057  & \texttt{revtex4-1}      & quant-ph \\
    2306.00207  & \texttt{amsart}         & math.AG \\
    2306.00417  & \texttt{revtex4-2}      & cond-mat.stat-mech \\
    2306.01308  & \texttt{revtex4}        & nucl-th \\
    2306.01691  & \texttt{acmart}         & cs.GR \\
    \bottomrule
  \end{tabular}
\end{table}

\subsection{Types of Inconsistencies}

\begin{table}[H]
  \caption{Number of inconsistencies found}
  \begin{tabular}{lrr}\toprule 
    & \multicolumn{2}{c}{Comparing \xetex with:} \\ \cmidrule(lr){2-3}
    Type of inconsistency & \pdftex & \luatex \\\midrule
    Text spacing        & 9   & 9    \\
    Line breaks         & 9   & 9    \\
    Images              & 3   & 4    \\
    Other formatting    & 3   & 2    \\
    Font formatting     & 2   & 0   \\
    Ligatures           & 2   & 0    \\
    Missing content     & 1   & 1    \\
    References          & 0   & 2    \\
    Number of pages     & 1   & 1    \\
    \bottomrule
  \end{tabular}
\end{table}

% -----------------------------------------------------------------------------

\section{Cross-engine Comparison}

\subsection{Inconsistencies in common document classes}

\begin{table}[H]
  \caption{Differences across common document classes (\xelatex vs \pdflatex)}
  \begin{tabular}{lrrrrrrr}\toprule 
    & \multicolumn{7}{c}{\% of papers with difference (\xelatex versus \pdflatex)} \\ \cmidrule(lr){2-8}
    Class (count)      & Missing styles & Missing content & Number of pages & Images & Text spacing & Line breaks & References  \\\midrule
    All compiled (342)  & 21.9  & 2.3   & 17.8  & 27.8  & 98.3  & 96.2  & 0.9  \\\midrule
    {article}	   (134)	& 17.2	& 3.7   &	37.3	& 41.0	& 97.0  & 97.7  & 1.5  \\
    {amsart}     (65)   & 6.2   & 0.0   & 21.5  & 21.5  & 96.8  & 88.9  & 0.0  \\
    {revtex4-2}  (43)   & 7.0   & 0.0   & 25.6  & 37.2  & 95.3  & 95.3  & 2.3  \\
    {IEEEtran}   (27)   & 63.0  & 0.0   & 63.0  & 70.4  & 100.0 & 88.9  & 0.0  \\
    {elsarticle} (26)   & 19.2  & 0.0   & 30.8  & 50.0  & 100.0 & 100.0 & 0.0  \\
    {revtex4-1}  (23)   & 0.0   & 0.0   & 30.4  & 52.2  & 100.0 & 100.0 & 0.0  \\
    {acmart}     (21)   & 19.0  & 0.0   & 28.6  & 52.4  & 95.0  & 95.0  & 0.0  \\
    {llncs}	     (8)    &	12.5  & 0.0   &	12.5  &	0.0   & 100.0	& 100.0 &	0.0 \\
    {mnras}	     (8)    &	25.0  & 0.0   &	25.0  &	25.0  & 87.5	& 100.0 &	0.0 \\
    {revtex4}    (7)    &	0.0   &	14.3  &	28.6  &	57.1  & 100.0	& 100.0 &	0.0 \\\bottomrule
  \end{tabular}
\end{table}

\begin{table}[H]
  \caption{Differences across common document classes (\xelatex vs \lualatex)}
  \begin{tabular}{lrrrrrrr}\toprule 
    & \multicolumn{7}{c}{\% of papers with difference (\xelatex versus \lualatex)} \\ \cmidrule(lr){2-8}
    Class (count)      & Missing styles & Missing content & Number of pages & Images & Text spacing & Line breaks & References  \\\midrule
    All compiled (340)  & 0.0   & 4.1   & 8.2   & 19.1  & 74.8  & 74.3  & 10.0 \\\midrule
    {amsart}     (65)   & 0.0   & 12.3  & 30.8  & 21.5  & 50.8  & 44.4  & 21.5 \\
    {revtex4-2}  (43)   & 0.0   & 0.0   & 23.3  & 37.2  & 74.4  & 76.7  & 4.7  \\
    {IEEEtran}   (27)   & 0.0   & 3.7   & 29.6  & 37.0  & 88.9  & 88.9  & 3.7  \\
    {elsarticle} (26)   & 0.0   & 0.0   & 23.1  & 38.5  & 80.0  & 80.0  & 0.0  \\
    {revtex4-1}  (23)   & 0.0   & 0.0   & 30.4  & 52.2  & 73.9  & 73.9  & 0.0  \\
    {acmart}     (21)   & 0.0   & 0.0   & 23.8  & 47.6  & 90.0  & 95.0  & 0.0  \\
    {llncs}	     (8)	  & 0.0	  & 0.0	  & 0.0	  & 0.0	  & 37.5	& 37.5  & 12.5 \\
    {mnras}	     (8)	  & 0.0	  & 0.0	  & 0.0	  & 12.5 	& 100.0	& 87.5  & 0.0  \\
    {revtex4}	   (7)	  & 0.0	  & 14.3 	& 28.6  & 57.1 	& 85.7	& 85.7  & 14.3 \\\bottomrule
  \end{tabular}
\end{table}

\clearpage
\subsection{Distribution of inconsistencies observed}

\begin{figure}[H]
  \Description{Venn diagram of the distribution of inconsistencies observed (\pdftex and \xetex)}
  \includegraphics[width=0.7\linewidth]{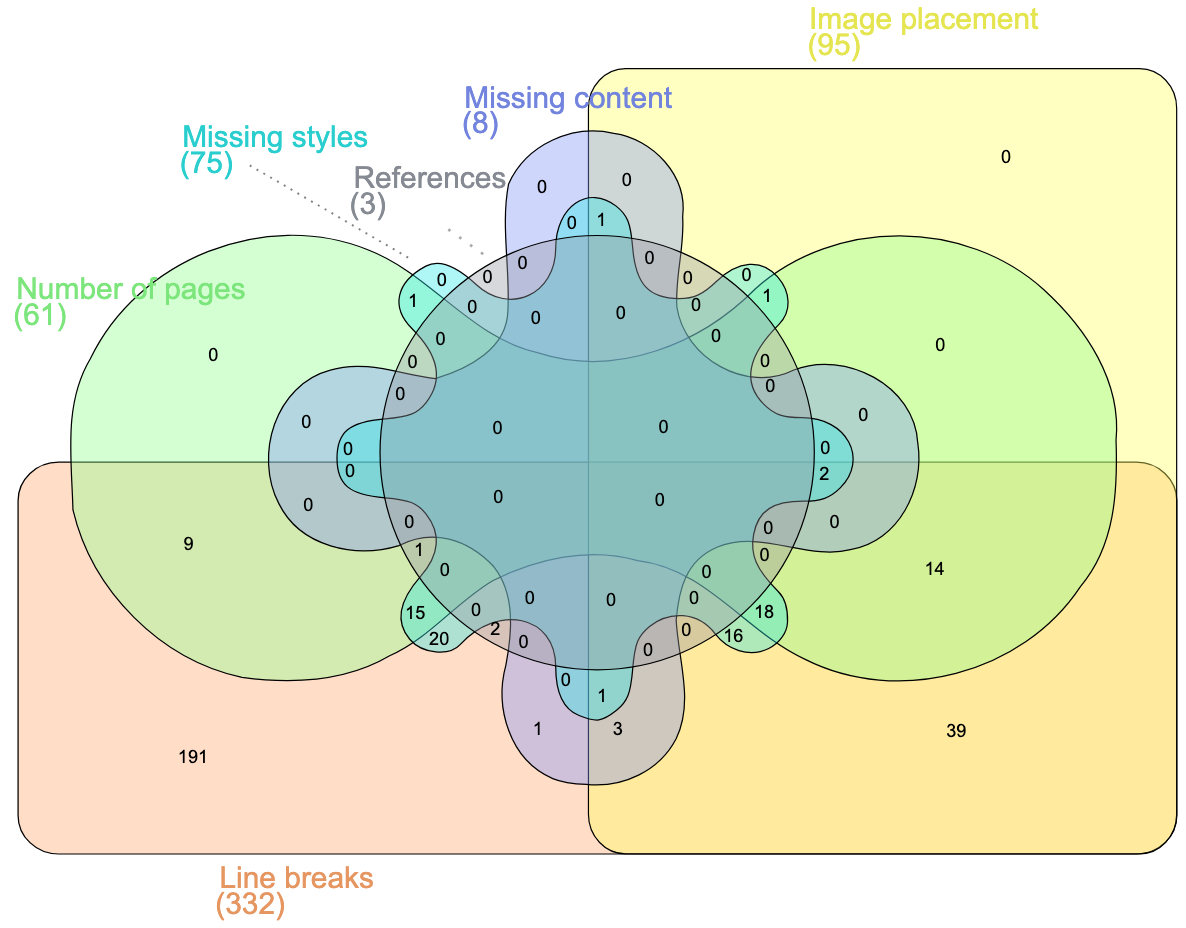} 
  \caption{Distribution of inconsistencies between \pdftex and \xetex}
\end{figure}

\begin{figure}[H]
  \Description{Venn diagram of the distribution of inconsistencies observed (\luatex and \xetex)}
  \includegraphics[width=0.7\linewidth]{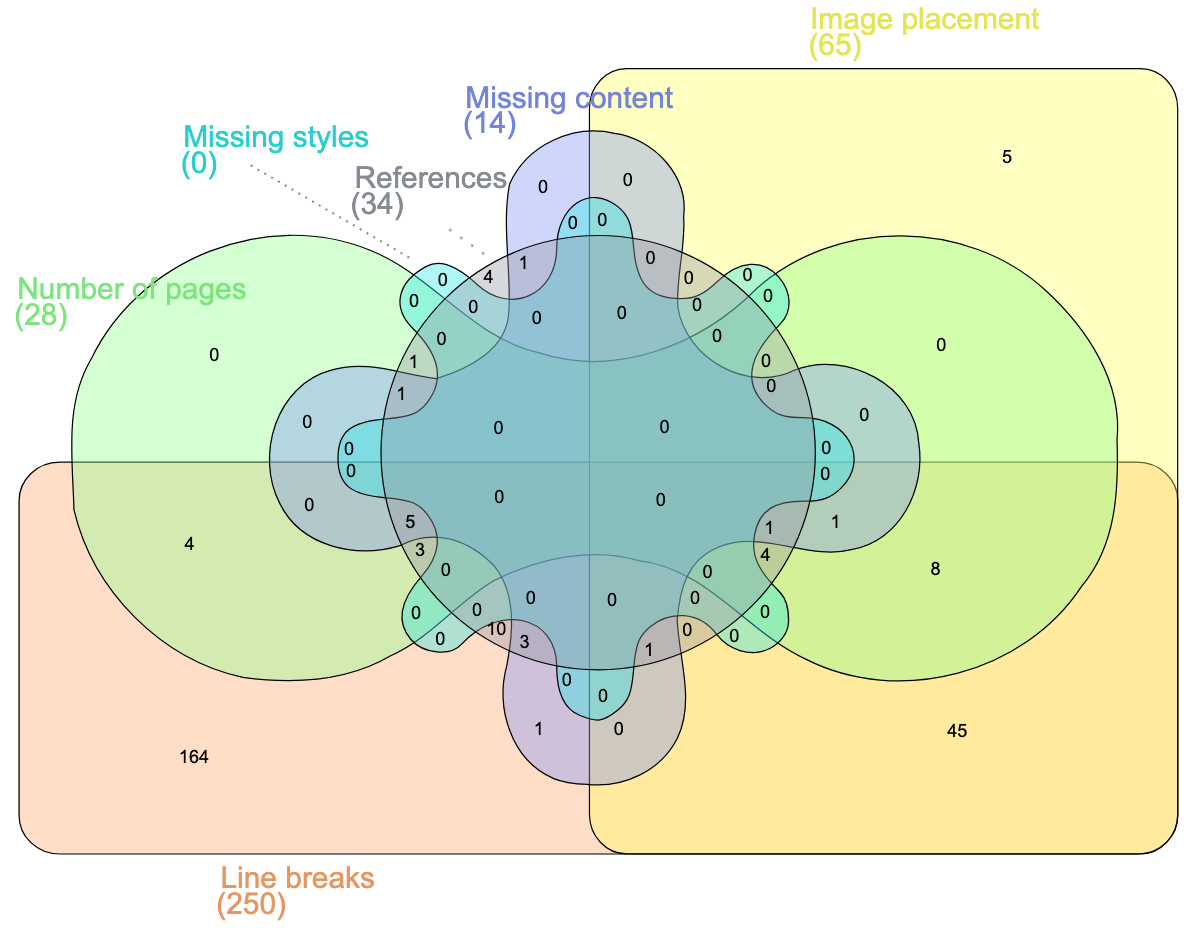} 
  \caption{Distribution of inconsistencies between \luatex and \xetex}
\end{figure}

% -----------------------------------------------------------------------------

\section{Cross-version Comparison}

\subsection{Inconsistencies in common document classes}

\begin{table}[H]
  \caption{Differences across common document classes (\texlive 2020 vs 2023)}
  \begin{tabular}{lrrrrrrr}\toprule 
    & \multicolumn{7}{c}{\% of papers with difference (\texlive 2020 versus 2023)} \\ \cmidrule(lr){2-8}
    Class (count)     & Missing styles & Missing content & Number of pages & Images & Text spacing & Line breaks & References  \\\midrule
    %                 sty    content pgs    imgs  textSPC lineSPC refs
    All compiled      & 5.6  & 6.5   & 7.9  & 1.6 & 51.2  & 14.8 & 11.6 \\\midrule 
    article     (133) & 5.3  & 3.0   & 6.8  & 3.0 & 52.6  & 9.8  & 7.5   \\ 
    amsart      (65)  & 0.0  & 1.5   & 0.0  & 1.5 & 52.3  & 3.1  & 3.1   \\ 
    revtex4-2   (43)  & 4.7  & 9.3   & 11.6 & 0.0 & 46.5  & 16.3 & 14.0  \\ 
    elsarticle  (26)  & 7.7  & 11.5  & 11.5 & 0.0 & 38.5  & 19.2 & 11.5  \\ 
    IEEEtran    (26)  & 0.0  & 15.4  & 7.7  & 0.0 & 34.6  & 19.2 & 15.4  \\ 
    revtex4-1   (23)  & 8.7  & 0.0   & 17.4 & 0.0 & 56.5  & 17.4 & 8.7   \\ 
    acmart      (21)  & 9.5  & 19.0  & 19.0 & 4.8 & 90.5  & 47.6 & 61.9  \\ 
    mnras       (8)   & 25.0 & 0.0   & 12.5 & 0.0 & 100.0 & 50.0 & 0.0   \\ 
    llncs       (8)   & 0.0  & 12.5  & 0.0  & 0.0 & 25.0  & 0.0  & 25.0  \\ 
    revtex4     (7)   & 0.0  & 0.0   & 0.0  & 0.0 & 14.3  & 0.0  & 0.0   \\ 
    \bottomrule
  \end{tabular}
\end{table}

\begin{table}[H]
  \caption{Differences across common document classes (\texlive 2020 vs 2021)}
  \begin{tabular}{lrrrrrrr}\toprule 
    & \multicolumn{7}{c}{\% of papers with difference (\texlive 2020 versus 2021)} \\ \cmidrule(lr){2-8}
    Class (count)     & Missing styles & Missing content & Number of pages & Images & Text spacing & Line breaks & References  \\\midrule
    All compiled      & 4.6  & 2.3   & 2.8  & 0.7 & 42.4  & 7.6  & 2.8 \\\midrule 
    article (133) & 5.3 & 0.8 & 3.0 & 1.5 & 43.6 & 4.5 & 1.5 \\ 
    amsart (65) & 1.5 & 0.0 & 0.0 & 0.0 & 50.8 & 1.5 & 0.0 \\ 
    revtex4-2 (43) & 4.7 & 0.0 & 0.0 & 0.0 & 44.2 & 4.7 & 0.0 \\ 
    elsarticle (26) & 7.7 & 3.8 & 3.8 & 0.0 & 26.9 & 7.7 & 3.8 \\ 
    IEEEtran (26) & 3.8 & 7.7 & 3.8 & 0.0 & 19.2 & 3.8 & 3.8 \\ 
    revtex4-1 (23) & 0.0 & 0.0 & 8.7 & 0.0 & 34.8 & 13.0 & 4.3 \\ 
    acmart (21) & 9.5 & 9.5 & 4.8 & 4.8 & 76.2 & 33.3 & 19.0 \\ 
    mnras (8) & 0.0 & 0.0 & 12.5 & 0.0 & 87.5 & 37.5 & 0.0 \\ 
    llncs (8) & 0.0 & 12.5 & 0.0 & 0.0 & 25.0 & 0.0 & 12.5 \\ 
    revtex4 (7) & 0.0 & 0.0 & 0.0 & 0.0 & 14.3 & 0.0 & 0.0 \\ 
    \bottomrule
  \end{tabular}
\end{table}

\begin{table}[H]
  \caption{Differences across common document classes (\texlive 2021 vs 2022)}
  \begin{tabular}{lrrrrrrr}\toprule 
    & \multicolumn{7}{c}{\% of papers with difference (\texlive 2021 versus 2022)} \\ \cmidrule(lr){2-8}
    Class (count)     & Missing styles & Missing content & Number of pages & Images & Text spacing & Line breaks & References  \\\midrule
    %                 sty    content pgs    imgs  textSPC lineSPC refs
    All compiled      & 1.4 & 	6.5 & 	5.3 & 	0.0 & 	15.5 & 	8.8 & 	10.0 \\\midrule 
    article     (133) & 0.8 & 	3.8 & 	3.8 & 	0.0 & 	12.8 & 	5.3 & 	6.8  \\ 
    amsart      (65)  & 0.0 & 	3.1 & 	0.0 & 	0.0 & 	4.6 & 	3.1 & 	3.1  \\ 
    revtex4-2   (43)  & 0.0 & 	7.0 & 	11.6 & 	0.0 & 	18.6 & 	16.3 & 	16.3 \\ 
    elsarticle  (26)  & 0.0 & 	7.7 & 	7.7 & 	0.0 & 	11.5 & 	7.7 & 	7.7  \\ 
    IEEEtran    (26)  & 0.0 & 	7.7 & 	7.7 & 	0.0 & 	23.1 & 	19.2 & 	15.4 \\ 
    revtex4-1   (23)  & 4.3 & 	8.7 & 	8.7 & 	0.0 & 	26.1 & 	13.0 & 	13.0 \\ 
    acmart      (21)  & 0.0 & 	14.3 & 	9.5 & 	0.0 & 	38.1 & 	19.0 & 	23.8 \\ 
    mnras       (8)   & 0.0 & 	0.0 & 	0.0 & 	0.0 & 	0.0 & 	0.0 & 	0.0  \\ 
    llncs       (8)   & 0.0 & 	12.5 & 	0.0 & 	0.0 & 	0.0 & 	0.0 & 	12.5 \\ 
    revtex4     (7)   & 0.0 & 	0.0 & 	0.0 & 	0.0 & 	0.0 & 	0.0 & 	0.0  \\ 
    \bottomrule
  \end{tabular}
\end{table}

\begin{table}[H]
  \caption{Differences across common document classes (\texlive 2022 vs 2023)}
  \begin{tabular}{lrrrrrrr}\toprule 
    & \multicolumn{7}{c}{\% of papers with difference (\texlive 2022 versus 2023)} \\ \cmidrule(lr){2-8}
    Class (count)     & Missing styles & Missing content & Number of pages & Images & Text spacing & Line breaks & References  \\\midrule
    %                 sty    content pgs    imgs  textSPC lineSPC refs
    All compiled      & 1.6  & 0.7  & 0.5 & 0.0 & 13.0  & 3.2  & 2.3  \\\midrule 
    article     (133) & 1.5  & 0.0  & 0.0 & 0.0 & 7.5   & 1.5  & 0.0  \\ 
    amsart      (65)  & 0.0  & 0.0  & 0.0 & 0.0 & 1.5   & 0.0  & 0.0  \\ 
    revtex4-2   (43)  & 0.0  & 0.0  & 0.0 & 0.0 & 20.9  & 2.3  & 0.0  \\ 
    elsarticle  (26)  & 0.0  & 0.0  & 0.0 & 0.0 & 7.7   & 0.0  & 0.0  \\ 
    IEEEtran    (26)  & 3.8  & 0.0  & 0.0 & 0.0 & 11.5  & 3.8  & 0.0  \\ 
    revtex4-1   (23)  & 0.0  & 0.0  & 4.3 & 0.0 & 13.0  & 0.0  & 0.0  \\ 
    acmart      (21)  & 0.0  & 14.3 & 4.8 & 0.0 & 61.9  & 47.6 & 47.6 \\ 
    mnras       (8)   & 37.5 & 0.0  & 0.0 & 0.0 & 100.0 & 0.0  & 0.0  \\ 
    llncs       (8)   & 0.0  & 0.0  & 0.0 & 0.0 & 12.5  & 0.0  & 0.0  \\ 
    revtex4     (7)   & 0.0  & 0.0  & 0.0 & 0.0 & 0.0   & 0.0  & 0.0  \\ 
    \bottomrule
  \end{tabular}
\end{table}

\subsection{Compilation success rates across different \texlive distributions}

\begin{table}[H]
  \caption{Comparison results (\%) from \texlive 2020 to 2023}
  \begin{tabular}{lrrr|r}\toprule 
  & \multicolumn{4}{c}{\texlive versions compared} \\ \cmidrule(lr){2-5}
    Comparison result /\%  & '20/'21 & '21/'22 & '22/'23 & '20/'23  \\\midrule
    Identical PDFs  & 20.4 & 80.1  & 85.2  & 30.3  \\
    Different PDFs  & 69.4 & 19.2  & 13.9  & 68.1  \\
    Compile failure & 1.2  & 0.7   & 0.9   & 1.6   \\\bottomrule
  \end{tabular}
\end{table}

\subsection{Pairwise comparison results over time}

\begin{table}[H]\centering
  \aboverulesep=0mm \belowrulesep=0mm
  \caption{Comparison results over time (2020 to 2023)}
  \yes\, identical output; \; \no\, different output \\
  \begin{tabular}{ccc|c|r}\toprule 
    \multicolumn{4}{c|}{Comparison results} & \multirow{2}{*}{Count} \\ \cmidrule{1-4}
    '20/'21 & '21/'22 & '22/'23 & '20/'23   \\\midrule
    \yes	& \yes	& \yes	& \yes	& 42.1\% \\
    \no 	& \yes	& \yes	& \no 	& 27.1\% \\
    \no 	& \no 	& \yes	& \no 	& 8.1\%  \\
    \yes	& \no 	& \yes	& \no 	& 7.9\%  \\
    \no 	& \yes	& \no 	& \no 	& 6.7\%  \\
    \yes	& \yes	& \no 	& \no 	& 3.0\%  \\
    \no 	& \no 	& \no 	& \no 	& 2.8\%  \\
    \no 	& \yes	& \no 	& \yes	& 1.2\%  \\
    \yes	& \no 	& \no 	& \no 	& 0.7\%  \\
    \yes	& \no 	& \no 	& \yes	& 0.5\%  \\\bottomrule
  \end{tabular}
\end{table}

\begin{table}[H]\centering
  \aboverulesep=0mm \belowrulesep=0mm
  \caption{Comparison results (\%) from \texlive 2020 to 2023}
  \label{tbl:versionCompareResuls}
  \begin{tabular}{lrrr|r}\toprule 
  & \multicolumn{4}{c}{\texlive versions compared} \\ \cmidrule(lr){2-5}
    Comparison result /\%  & '20/'21 & '21/'22 & '22/'23 & '20/'23  \\\midrule
    Identical PDFs  & 20.4 & 80.1  & 85.2  & 30.3  \\
    Different PDFs  & 69.4 & 19.2  & 13.9  & 68.1  \\
    Compile failure & 1.2  & 0.7   & 0.9   & 1.6   \\\bottomrule
  \end{tabular}
\end{table}

% -----------------------------------------------------------------------------

\section{Root Cause Analysis}

\subsection{Sampled Documents}

In \textbf{RQ3}, we studied 26 documents:
\begin{itemize}
  \item 7 compile failures (4 root causes)
  \item 6 reversions (3 root causes)
  \item 13 changes from \texlive 2022 to \texlive 2023 (3 root causes)
\end{itemize}

\begin{table}[H]
  \caption{Root causes of compile failures}
  \begin{tabular}{llll}\toprule 
    \textbf{arXiv ID} & \textbf{Years with compilation failures} & \textbf{Triage} & \textbf{Root cause} \\\midrule
    2306.00003 & 2023                    & Reported bug       & \texttt{jmlr} package \\
    2306.05750 & 2020                    & Fixed bug          & Adding a linebreak in footnotes \\
    2306.03822 & 2020                    & Expected behaviour & Imported (\texttt{tabularray.sty}) which did not exist yet \\
    2306.00275 & 2020                    & Expected behaviour & Imported (\texttt{tabularray.sty}) which did not exist yet \\
    2306.00490 & 2020, 2021, 2022, 2023  & Expected behaviour & Syntax error by author \\
    2306.00055 & 2020, 2021, 2022, 2023  & Expected behaviour & Syntax error by author \\
    2306.00030 & 2020, 2021, 2022, 2023  & Expected behaviour & Syntax error by author \\
    \bottomrule
  \end{tabular}
\end{table}

\begin{table}[H]
  \caption{Root causes of reversions}
  \begin{tabular}{lrrll}\toprule 
    \textbf{arXiv ID} & \textbf{Change introduced} & \textbf{Change reverted} & \textbf{Triage} & \textbf{Root cause} \\\midrule
    2306.00001 &  2021  &  2023  & Fixed bug & \texttt{siunitx} package applied font styles inconsistently \\
    2306.01403 &  2021  &  2023  & Fixed bug & \texttt{revtex4-2} package did not detect the \texttt{eqnarray} environment \\
    2306.03237 &  2021  &  2023  & Fixed bug & \texttt{revtex4-2} package did not detect the \texttt{eqnarray} environment \\
    2306.00365 &  2021  &  2023  & Fixed bug & \texttt{revtex4-2} package did not detect the \texttt{eqnarray} environment \\
    2306.00746 &  2022  &  2023  & Fixed bug & Handling of newlines after the \verb|\eqno| macro \\
    2306.01235 &  2022  &  2023  & Fixed bug & Handling of newlines after the \verb|\eqno| macro \\
    \bottomrule
  \end{tabular}
\end{table}

\begin{table}[H]
  \caption{Root causes of changes only in \texlive 2022 to 2023}
  \begin{tabular}{llll}\toprule 
    \textbf{arXiv ID} & \textbf{Inconsistency} & \textbf{Triage} & \textbf{Root cause} \\\midrule
    2306.00285 & Vertical spacing                  & Fixed bug          & Importing \texttt{hyperref} package changes line spacing \\
    2306.00052 & Importing \texttt{colortbl} changes horizontal spacing & Confirmed bug      & Inconsistent behaviour in \latex \texttt{array} package \\
    2306.00329 & Spacing with alignment characters & Expected behaviour & Redefining tilde character in core \latex package \\
    2306.00415 & Spacing with alignment characters & Expected behaviour & Redefining tilde character in core \latex package \\
    2306.01017 & Spacing with alignment characters & Expected behaviour & Redefining tilde character in core \latex package \\
    2306.00372 & Spacing with alignment characters & Expected behaviour & Redefining tilde character in core \latex package \\
    2306.00060 & Spacing with alignment characters & Expected behaviour & Redefining tilde character in core \latex package \\
    2306.01838 & Spacing with alignment characters & Expected behaviour & Redefining tilde character in core \latex package \\
    2306.00269 & Spacing with alignment characters & Expected behaviour & Redefining tilde character in core \latex package \\
    2306.00510 & Spacing with alignment characters & Expected behaviour & Redefining tilde character in core \latex package \\
    2306.00537 & Spacing with alignment characters & Expected behaviour & Redefining tilde character in core \latex package \\
    2306.00469 & Spacing with alignment characters & Expected behaviour & Redefining tilde character in core \latex package \\
    2306.01515 & Spacing with alignment characters & Expected behaviour & Redefining tilde character in core \latex package \\
    \bottomrule
  \end{tabular}
\end{table}

% -----------------------------------------------------------------------------

\section{Comparison Methods}

The full source code of our tool, including the implementation of all comparison methods, is available at 
\url{https://github.com/jovyntls/inconsistencies-in-tex}, or \url{https://zenodo.org/records/12669708}.
The following is a summary of the tools used in performing comparisons:

\begin{itemize}
  \item \texttt{diff-pdf} for pixel-wise comparisons of PDFs. Available at: \url{https://github.com/vslavik/diff-pdf}
  \item \texttt{PyMuPDF} for extracting text and fonts from PDFs. Available at: \url{https://pypi.org/project/PyMuPDF/}
  \item \texttt{opencv} for feature extraction. Available at: \url{https://opencv.org/get-started/.} Specifically, we used the:
    \begin{itemize}
      \item Structural Similarity (SSIM) Index
      \item Complex Wavelet Structural Similarity (CW-SSIM) Index
      \item Scale-Invariant Feature Transform (SIFT) algorithm
      \item Oriented FAST and Rotated BRIEF (ORB) algorithm
    \end{itemize}
\end{itemize}

%% file: sections/abstract.tex
\tex is a widely used typesetting system adopted by most publishers and professional societies.
While \tex is responsible for generating a significant number of documents, irregularities in the \tex ecosystem may produce inconsistent documents.
These inconsistencies may occur across different \tex engines or different versions of \tex distributions, resulting in failures to adhere to formatting specifications, or the same document rendering differently for different authors.
In this work, we investigate and quantify the robustness of the \tex ecosystem through a large-scale study of 432 documents.
We developed an automated pipeline to evaluate the cross-engine and cross-version compatibility of the \tex ecosystem.
We found significant inconsistencies in the outputs of different \tex engines: only 0.2\% of documents compiled to identical output with \xetex and \pdftex due to a lack of cross-engine support in popular \latex packages and classes used in academic conferences.
A smaller---yet significant---extent of inconsistencies was found across different \texlive distributions, with only 42.1\% of documents producing the same output from 2020 to 2023.
Our automated pipeline additionally reduces the human effort in bug-finding:
from a sample of 10 unique root causes of inconsistencies, we identified two new bugs in \latex packages and five existing bugs that were fixed independently of this study.
We also observed potentially unintended inconsistencies across different \texlive distributions beyond the updates listed in changelogs.
We expect that this study will help authors of \tex documents to avoid unexpected outcomes by understanding how they may be affected by the often undocumented subtleties of the \tex ecosystem, while benefiting developers by demonstrating how different implementations result in unintended inconsistencies.

%% file: sections/introduction.tex
\section{Introduction}

\tex is a typesetting system widely used to produce output of the highest typographic quality. 
\latex is a set of macros built on top of \tex; \latex systems are currently the dominant typesetting environment used in scientific publishing~\cite{Gaudeul:latexPopularity}, and have been adopted by most publishers and professional societies as their preferred format~\cite{CTAN:texAndFriends, Brischoux:DontFormatManuscripts}.
Overleaf, an online \latex editor, amassed 11 million users in 2022~\cite{Overleaf:Users}.
Evidently, the \tex ecosystem is responsible for generating a large amount of documents.

Bugs and inconsistencies in the \tex ecosystem may produce documents that are incorrect or inconsistent, potentially affecting the communication of information.
Given a source \tex file, the output of each \tex engine is usually expected to be visually identical. 
Unexpected differences in output between different engines compromise the quality of typeset documents.
Examples of such differences include information loss, incorrect formatting, or inconsistent typography.
Differences in output are highly undesirable when preparing documents that must adhere to a strict specification, such as in academic conferences or journals where violating formatting requirements results in desk rejection.

While authors may expect the compiled output of their \tex files to remain consistent across different \tex engines and different versions of \texlive, this hypothesis has not been systematically studied.
To our knowledge, there is currently no existing work on a systematic comparison of the output produced by different \tex engines and \texlive distributions.
While release notes and documentation are helpful, dependencies on various \latex macros and simultaneous updates of different packages make it difficult for authors to determine how they might be affected by different engines or \texlive versions respectively.

% The closest related work is on testing compilers, such as automated differential testing on C compilers~\cite{Yang:FindingBugsCCompilers}.
% While this work similarly aims to use an automated differential testing approach in bug-finding, the differences in the output of a C compiler (a well-defined program output) and a \tex engine (where a range of different outputs may be considered acceptable) presented additional challenges in this work.
% Another related work is Donaldson et al.'s work on metamorphic testing for graphics shade compilers~\cite{Donaldson:TestingGraphicsShaderCompilers}. 
% Similar to our work, this paper presents an approach for testing a compiler which allows for some variance in the output, by using chi-squared distance algorithms to compare image histograms and identify meaningful differences from non-meaningful ones.
% Such image-based approaches were not suitable in our work as minor differences in line breaks and hyphenations resulted in noise in image histograms, which obscured other meaningful differences.

Motivated by the limited information available for authors to navigate the complex \tex ecosystem, we pose and answer the following research questions (RQs):

\begin{enumerate}[label*=\textbf{RQ\arabic*}, leftmargin=0pt, itemsep=5pt, topsep=5pt, itemindent=\widthof{\textbf{RQ00}}]
  \item \textit{How interchangeable are different \tex engines?} \label{rq1:interengine}

    Knowing this helps authors make informed choices between \tex engines;
    understand how their choice of engine may affect their writing experience and output;
    and avoid unexpected outcomes when their documents are compiled with different engines.
    While each \tex engine may list its unique features, implementation differences might introduce unintended inconsistencies that are of interest to authors.
    Since the PDFs produced by different engines may inevitably contain minor differences that are transparent to end users (such as in differences in metadata), we consider outputs \textit{consistent} if they are visually identical.
    While different \tex engines are expected to produce consistent output when given the same input file, we hypothesise that this may not always be the case due to differences in their implementations.

  \item \sloppy \textit{How interchangeable are different versions of \texlive?} \label{rq2:interversion}

    Knowing this helps authors of \tex files to understand the potential consequences of upgrading---or failing to upgrade---their \tex distributions.
    Characterising the differences observed in this study concretises the inconsistencies that are of interest to authors.
    Authors may thus make an informed decision on when to upgrade their \tex-related software or which versions to use to ensure that their documents meet given specifications.

  \item \textit{Are inconsistencies across \texlive versions considered bugs?} \label{rq3:bugs}

    Knowing this helps developers of \tex-related software (like \latex packages) to understand how changes in behaviour are triaged and what extent of inconsistency is considered reasonable between versions.
    Authors may also benefit by knowing what to expect from, and how they may be affected by, updating their \texlive distribution.
    We chose to focus on \texlive versions over \tex engines as the subtle differences between engines and the lack of a formal specification make it difficult to define expected behaviour in each engine.

\end{enumerate}

To answer these questions, we conducted a large-scale empirical study on the inconsistencies in the \tex ecosystem. 
We studied the differences between the outputs of three different \tex engines as well as four different versions of \texlive distributions when compiling the same source files from 432 different documents, automating the comparison to identify common differences between engines.
We additionally used various metrics such as pixel-wise differences and textual differences to measure ``interchangeability'' by characterising these differences in a manner accessible for authors and users of the \tex ecosystem.
In summary, we found:

\begin{itemize}
  \item Several incompatibilities of \latex classes with different \tex engines. 
    These incompatibilities range from compilation errors to significant differences in the output PDF.
    Several popular \latex classes, such as document classes for conference submissions, were compatible only with \pdftex and incompatible with other engines.

  \item A large extent of differences between \tex engines: only 0.2\% of documents had identical (pixel-wise) output with \pdftex and \xetex, and only 1.4\% had identical output with \luatex and \xetex.

  \item A smaller, but still significant, extent of differences between different \texlive distributions: only 42.1\% of documents compiled to the same output across all four releases. 
    The change from \texlive 2020 to 2021 was the most significant, with 27.1\% of documents producing different output across these releases but consistent output in subsequent releases.

  \item Two new bugs, and five existing bugs, in \latex packages and classes.
    Other inconsistencies found were considered expected differences, with varying levels of documentation.

  \item Undocumented changes in the output of popular document classes over different versions of \texlive, highlighting the importance of verifying the \texlive version used in compiling documents that must adhere to strict specifications.

  \item Minor differences (such as the spacing between letters or words differing by a few pixels) were common. 
    Although these inconsistencies might not affect the information conveyed in the document, multiple inconsistencies may culminate in larger and significant differences (such as a different number of pages in the output document).
\end{itemize}

All data, scripts, and results are available at 
\url{https://zenodo.org/records/12669708}.
To our knowledge, this large-scale study is the first such study on the inconsistencies of the \tex ecosystem. 
By systematically quantifying the large extent of inconsistencies between different versions of \tex Live and \tex engines, we help authors understand how to avoid unexpected outcomes, especially in collaborative cross-platform or cross-version contexts.
Additionally, users of the \tex ecosystem may gain a better understanding of which \tex engine will best suit their needs, and how the \tex engine they choose to use may affect their output document. 
By highlighting the undocumented changes in the output of popular document classes and packages over different distributions and engines, we help developers within the \tex ecosystem (such as developers of \latex packages) gain a better awareness of how to develop cross-compatible packages to expand their user base.
The inconsistencies we have uncovered provide a strong starting point for developing test cases under more controlled environments.
Finally, our automated pipeline reduces the human effort in finding bugs in \tex by identifying documents that are likely to trigger bugs. We used it to find seven bugs, two of which were previously unknown ones.

%% file: sections/background.tex
\section{Background}

\paragraph{\texlive} 

\texlive is a software distribution that provides a comprehensive \tex system including various packages, macros, and engines~\cite{TeXLive:TeXLive}.
\texlive is released yearly, and while it is possible to update a \tex distribution between official releases, it is \emph{``not necessary or particularly recommended''} especially since the set of packages and programs in the official release is \emph{``the only one which gets tested as a coherent release''}~\cite{TeXLive:Availability}.

\paragraph{\tex engines}

Since \tex's release in 1978, other engines have been developed to extend the functionality of the original \tex program. 
This paper discusses three of these engines, all derived from \tex:

\begin{enumerate}
  \item \sloppy \pdftex, and its \latex variant \pdflatex, is a backward-compatible extension of \tex that can directly output PDF (``Portable Document Format'') files from \tex files~\cite{pdfTeXmanual}.
  \item \xetex is a Unicode \tex engine that enables convenient use of OpenType fonts and multilingual typesetting. It is not backward-compatible with \tex~\cite{XeTeXIntro}.
  \item \luatex introduces a Lua interpreter, allowing authors to integrate Lua code into their macros. It is not backward-compatible with \tex~\cite{LuaTeXmanual}.
\end{enumerate}

\paragraph{PDF outputs}

Given an input file, \tex aims to be system-agnostic in generating output: input files should \emph{``generate the same output on any system on which they were processed---same hyphenation, same line breaks, same page breaks, same everything''}~\cite{TUG:texHistory}.
While different \tex engines are not required to produce the same output given the same input file,
these three engines are compatible with the \latex format, and thus should produce relatively similar output according to this format.
Engines like \luatex \emph{``may be used as a drop-in replacement''} most of the time; for both \luatex and \xetex, due to the interfaces provided by the \latex format, \emph{``for most LaTeX end users, these subtleties are transparent''}~\cite{TexFaq:xetexLuatex}.

% \paragraph{Comparing outputs}

% Since the outputs of different engines are not expected to be exactly identical given the same input file, a direct pixel-wise comparison would be too coarse-grained as an approach to identify meaningful differences.
% Instead, we aimed to characterise the types of differences observed. 
% This presents a challenge in comparing outputs of different \tex engines and identifying the various types of inconsistencies present.

%% file: sections/motivating-study.tex
\section{Motivating Study}

As a motivating study to inform the design of our experimental methodology, we investigate different methods of comparing documents compiled from different \tex engines to determine how outputs differ between \tex engines.
We additionally identify and characterise some typical types of differences observed.

\subsection{Methodology}

We obtained 10 \tex source files from arXiv.org~\cite{arXiv:home}, an online open-access archive of scholarly articles, across different taxonomies and document classes (refer to the supplementary materials for a detailed list).
Each \tex file was compiled with \pdftex, \xetex and \luatex, yielding three output PDFs (one from each engine). 

We used four different approaches to pairwise comparisons of PDFs.
First, we performed pixel-by-pixel comparisons between PDFs produced by different engines.
Next, we used text- and font-based comparisons. % on the textual content, and font-based comparisons on the fonts extracted.
Lastly, we used features-based comparisons. We converted the first 3 and last 3 pages of each PDF to an image and used the four common computer vision algorithms (namely SIFT, ORB, SSIM, and CWSSIM) in \texttt{opencv} to identify similarities between significant structures or regions in each image.

Finally, we studied the differences as indicated by these comparison methods, identified their root causes, and classified them based on their root causes.
Both authors then discussed and agreed on the identified root causes to reduce subjectivity and bias in this step.

\subsection{Findings}

All 10 sources compiled successfully with \pdftex and \luatex.
One source failed to compile with \xetex due to the \tex file importing the \texttt{inputenc} package which explicitly throws an error to halt compilation and inform authors that the package is not supported by \xetex and \luatex.
This compilation failure was thus expected behaviour and not a bug in any engine.
Consequently, we observed that while engines aim to produce relatively similar output, engine-specific \latex packages might result in incompatibilities.

Of the 9 sources that successfully compiled a PDF with all three \tex engines, we found differences in all pairwise comparisons between PDFs compiled from the same source on different engines.
These differences ranged from inconsistencies in formatting, ligatures (a typographical feature where letters are joined to improve legibility), and content.
The supplementary materials provide detailed statistics on these differences.
In the rest of this section, we detail the most common types of inconsistencies observed.

\paragraph{Differences in text spacing} 

\begin{figure}[tb]
  \Description[Different line breaks in LuaTeX and XeTeX]{Example of LuaTeX and XeTeX compiling the same input document with different line breaks}
  \centering
  \setlength{\fboxsep}{0pt}%
  \begin{subfigure}{.75\columnwidth}
    \fbox{\includegraphics[width=0.98\linewidth]{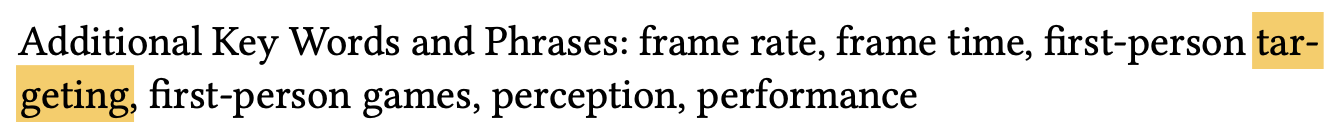}}
    \caption{\xelatex output}
    \label{fig:sampleDiff:wordBreaksXetex}
  \end{subfigure}
  \begin{subfigure}{.75\columnwidth}
    \fbox{\includegraphics[width=0.98\linewidth]{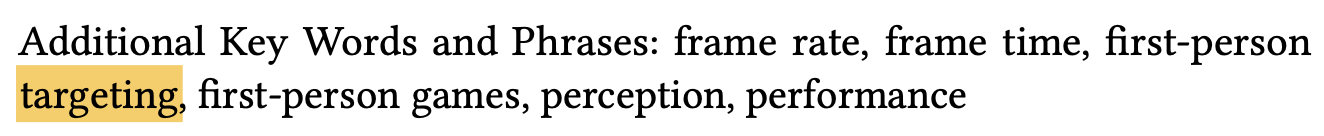}}
    \caption{\lualatex output}
    \label{fig:sampleDiff:wordBreaksLuatex}
  \end{subfigure}
  \caption{Differences in hyphenation in arXiv:2306.01691~\cite{arxiv:2306.01691}}
  \label{fig:sampleDiff:wordBreaks}
\end{figure}

The most common type of inconsistency we observed was differences in text spacing, which occurred in all 18 pairwise comparisons.
This is illustrated in Figure~\ref{fig:sampleDiff:wordBreaks}, where \xetex and \luatex output the same paragraph of text with a line break at different points due to very small differences in the size of---and space between---individual characters.
While this may seem like a small inconsistency, in one document, accumulated differences in text spacing and line breaks resulted in a different number of pages when compiled with \pdftex compared to \xetex despite having no differences in content.

\paragraph{Differences in formatting} 

\begin{figure}[tb]
  \Description[Different font styles in PDFTeX and XeTeX]{Example of PDFTeX and XeTeX compiling the same input document with different font styles}
  \centering
  \setlength{\fboxsep}{0pt}%
  \begin{subfigure}{.75\columnwidth}
    \centering
    \fbox{\includegraphics[width=0.98\linewidth]{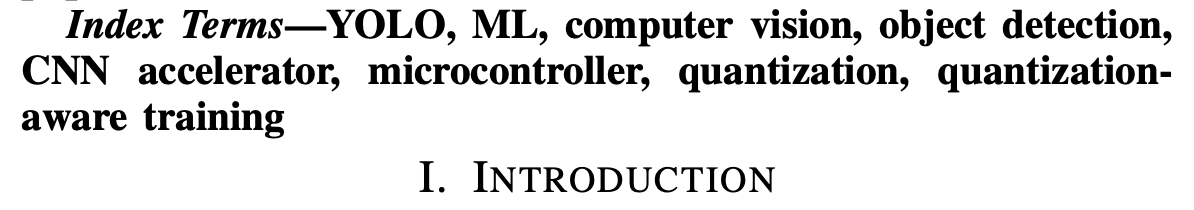}}
    \caption{\pdflatex output}
    \label{fig:sampleDiff:missingStylesPdftex01}
  \end{subfigure}
  \begin{subfigure}{.75\columnwidth}
    \centering
    \fbox{\includegraphics[width=0.98\linewidth]{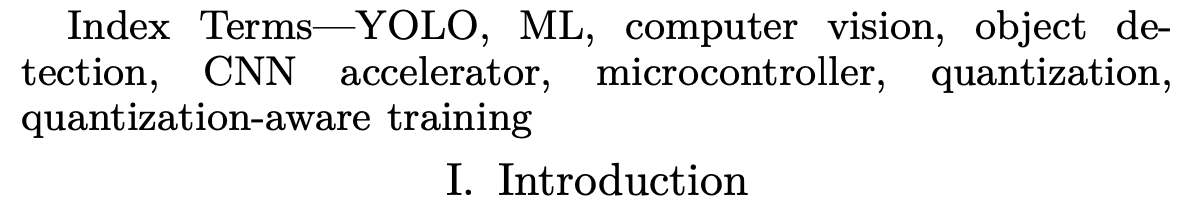}}
    \caption{\xelatex output}
    \label{fig:sampleDiff:missingStylesXetex01}
  \end{subfigure}
  \caption{Differences in font styles in the \texttt{IEEEtran} class~\cite{arxiv:2306.00001}}
  \label{fig:sampleDiff:missingStyles01}
\end{figure}

Formatting inconsistencies occurred between \pdftex and \xetex for two sources.
Figure~\ref{fig:sampleDiff:missingStyles01} shows an inconsistency in formatting between the documents produced by \pdftex (Figure~\ref{fig:sampleDiff:missingStylesPdftex01}) and \xetex (Figure~\ref{fig:sampleDiff:missingStylesXetex01}), where font styles (specifically \textbf{bold text}, \textit{italics} and \textsc{smallCaps}) failed to render in \xetex.

\paragraph{Differences in content} 

\begin{figure}[tb]
  \Description[Different header content in PDFTeX and XeTeX]{Example of PDFTeX and XeTeX compiling the same input document with different header content}
  \centering
  \setlength{\fboxsep}{0pt}%
  \begin{subfigure}{.7\columnwidth}
    \centering
    \fbox{\includegraphics[width=0.90\linewidth]{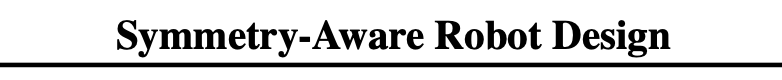}}
    \caption{\pdflatex output}
    \label{fig:sampleDiff:headerContentPdftex}
  \end{subfigure}
  \begin{subfigure}{.7\columnwidth}
    \centering
    \fbox{\includegraphics[width=0.90\linewidth]{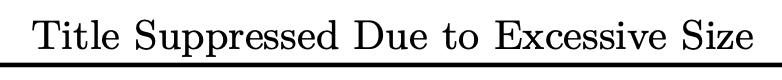}}
    \caption{\xelatex output}
    \label{fig:sampleDiff:headerContentXetex}
  \end{subfigure}
  \caption{Different page header text for arXiv:2306.00036~\cite{arxiv:2306.00036} using the \texttt{icml2023} class (for ICML-2023)}
  \label{fig:sampleDiff:headerContent}
\end{figure}

Differences in content occurred between \xetex and \pdftex for one source, and between \xetex and \luatex for another source.
Figure~\ref{fig:sampleDiff:headerContent} shows an example of this inconsistency, where \xetex and \pdftex rendered different text for a page header.
This difference was caused by a hard-coded maximum line height in the \texttt{icml2023} document class, which was exceeded in the document compiled by \xetex (Figure~\ref{fig:sampleDiff:headerContentXetex}) due to differences in the font used.

\subsection{Effectiveness of Comparison Methods}

We found identifying inconsistencies to be a significant challenge.
Due to the complexity of PDF output and the visual basis of determining similarity, we found that comparing PDFs---and identifying different types of inconsistencies between them---posed a key challenge in our initial study.
In the rest of this section, we describe how each comparison method performed, and how our findings using each comparison method informed our eventual methodology.

\paragraph{Pixel-based} 

While the naive method of using pixel-based comparisons could detect whether two documents were identical, it failed to differentiate between different types of inconsistencies.
Since every document had differences in text spacing, we found this comparison method insufficient as it flagged every comparison as being non-identical but did not help to identify more ``interesting'' inconsistencies such as missing content or differences in formatting.

\paragraph{Text-based} 

The text-based method of comparison could identify differences in content that were otherwise hard to detect, which helped to prevent false negatives (\emph{i.e.}, overlooked inconsistencies).
However, it failed to detect differences in formatting or other visual differences outside of the text content, making it unviable as a standalone comparison method.
Additionally, text-based comparisons were prone to false positives (\emph{i.e.}, false alarms due to incorrectly-flagged differences), as visually similar text was often extracted differently due to different handling of font and encoding across different engines.
For instance, to overcome the limited number of characters in OT1, \pdftex renders accented characters by combining a letter with an accent. The ``\'e'' character (the letter ``e'' with an acute diacritic) would be extracted as ``\'{}e'' in a PDF generated by \pdftex, and ``\'e'' in a PDF generated by \luatex or \xetex, despite being correctly rendered in all three engines.
% To reduce the rate of false positives, we applied an additional transformation step in our final methodology to filter out inconsistencies that resulted from different encoding.
% We additionally refined the text extraction process to ignore other differences that would be transparent to users, such as differences in how whitespace is encoded.

\paragraph{Font-based} 
Font-based comparison methods are a finer-grained tool to detect formatting inconsistencies. 
Due to the different font encodings, it was common for documents compiled by different engines to look identical, but use completely different fonts, which resulted in false positives.
By comparing the number of fonts instead of performing an exact match, we detected both of the formatting inconsistencies identified in this motivating study.

\paragraph{Feature-based} 

% \sloppy{} While computer-vision algorithms performed well on images with minimal content, we found that they failed to cope with entire pages of text.
% Due to the large amount of text in each image, it was not computationally feasible to consider all data present in the image, necessitating a trade-off between the accuracy and the running time of the algorithm.
% However, reducing the running time by reducing the number of features resulted in high rates of both false positives and false negatives.
% Feature-based comparisons were thus ineffective in our initial study.
% XXX: overfull hbox, but \sloppy{} increases the page count
While computer-vision algorithms performed well on images with minimal content, we found that they failed to cope with entire pages of text.
Due to the large amount of text, it was computationally infeasible to consider all data present in the image, necessitating trade-offs between the algorithm's accuracy and run time.
However, reducing run time by reducing the number of features resulted in high false positive and false negative rates.
Feature-based comparisons were thus ineffective in our initial study.

\paragraph{Summary} 

In this initial study, we found that 
pixel-based comparison methods detected the presence of inconsistencies in all comparisons performed, but were unable to characterise the inconsistencies found, which diminished its effectiveness as it failed to identify ``interesting'' differences.
Text-based comparison methods detected subtle differences in content, but had high false positive rates due to different encoding formats, and could not detect formatting inconsistencies.
Font-based comparisons could identify differences in text formatting, but a direct comparison of fonts was not feasible due to large differences in font encodings between \tex engines.
Finally, feature-based comparisons were not computationally feasible due to the complexity of the output documents.

Based on these findings, we developed a more refined comparison method using a combination of text-, font-, feature-, and pixel-wise methods to identify inconsistencies in PDF documents.
These are further detailed in Section~\ref{sec:methodology:characterisingDifferences}.

% \begin{figure}[tb]
%   \centering
%   \setlength{\fboxsep}{0pt}%
%   \begin{subfigure}{.8\columnwidth}
%     \fbox{\includegraphics[width=0.98\linewidth]{sample-diff-missing-styles-pdflatex-2306-00027.png}}
%     \caption{\pdflatex output}
%     \label{fig:sampleDiff:missingStylesPdftex}
%   \end{subfigure}
%   \begin{subfigure}{.8\columnwidth}
%     \fbox{\includegraphics[width=0.98\linewidth]{sample-diff-missing-styles-xelatex-2306-00027.png}}
%     \caption{\xelatex output}
%     \label{fig:sampleDiff:missingStylesXetex}
%   \end{subfigure}
%   \caption{Differences in styles (\textsc{smallCaps} and \textit{italics} missing) in arxiv:2306.00027~\cite{arxiv:2306.00027} using the \texttt{IEEEtran} class}
%   \label{fig:sampleDiff:missingStyles}
% \end{figure}

% However, we considered most of the differences highlighted by \texttt{diff-pdf} to be acceptable differences. 
% Figure~\ref{fig:sampleOutputDiffFontSpacing} shows an example of a differences between the output produced by \xetex (Figure~\ref{fig:sampleOutputDiffFontSpacingXetex}) and \pdftex (Figure~\ref{fig:sampleOutputDiffFontSpacingPdftex}). 
% When overlaying both PDFs, we can see that the positions of text differ by just a few pixels (Figure~\ref{fig:sampleOutputDiffFontSpacingDiff}).

%% file: sections/methodology.tex
\section{Methodology}

To answer our research questions, we compared and analysed the compilation outputs of a large number of \tex files.
To maximise the efficiency of analysis, we designed and implemented an automated pipeline using a comparative approach to identify inconsistencies.

\subsection{Dataset} 

To obtain a large amount of \tex source files, we extracted papers with downloadable source code from \url{arxiv.org}.
We chose to use arXiv as it presents a high volume and wide variety of papers, most of which have a \tex source. 
As many of these sources contain multiple files, we identified compilation entry points using regular expressions to simulate a fuzzy search for common file names (\emph{e.g.}, ``main.tex'', ``manuscript.tex'') and keywords (\emph{e.g.}, \texttt{documentclass}), similar to the approach used by arXiv~\cite{AutoTeX}.

For each of the 155 taxonomies in arXiv~\cite{arXiv:CategoryTaxonomy}, we queried 3 papers published in June 2023 (arbitrarily chosen) for a target total of 455 papers across a variety of subjects. 
By covering each taxonomy, we aimed to diversify the testing inputs across different usages of \tex, document classes, and packages.
This approach yielded 432 papers in total as some taxonomies had fewer than 3 papers published in that month with \tex sources.

\subsection{Measuring Interchangeability}

To investigate the interchangeability of different \tex engines (\ref{rq1:interengine}) and different versions of \texlive (\ref{rq2:interversion}), we developed an automated pipeline for compiling \tex sources and comparing output PDFs as shown in Figure~\ref{fig:approachOverview}.

\tikzset{
  process/.style = {ellipse, minimum width=0.6cm, inner sep=1pt, text centered, draw=black, fill=green!30, shape aspect=1.8},
  file/.style = {rectangle, minimum width=0.3cm, minimum height=0.4cm, text centered, draw=black, fill=blue!30},
  longfile/.style = {rectangle, minimum width=0.5cm, minimum height=0.5cm, text centered, draw=black, fill=blue!30, text width=1.2cm, inner sep=2pt},
  source/.style = {cylinder, minimum width=1cm, minimum height=0.5cm, text centered, draw=black, fill=gray!30, shape border rotate=90, shape aspect=0.5},
  misc/.style = {cloud, text centered, draw=black, fill=yellow!30, inner sep=0pt, text width=1.2cm, shape aspect=1.8},
  arrow/.style = {thick,->,>=stealth}
}

\begin{figure}[tb]
  \Description{An overview of the implementation of the automated pipeline}
  \centering
  \begin{tikzpicture}[node distance=1.2cm, font=\footnotesize]
    \node (compressedArxivSrc) [longfile] {Compressed source};
    \node (extractedArxivSrc) [longfile, below of=compressedArxivSrc, yshift=1mm] {Extracted source};
    \node (cleanedTexFiles) [longfile, below of=extractedArxivSrc, yshift=1mm] {``Cleaned'' \tex files};
    \node (texEntrypoint) [longfile, below of=cleanedTexFiles, yshift=1mm] {\tex entrypoint};

    \node (arxivSource) [source, right of=compressedArxivSrc, xshift=1cm] {arXiv};

    \node (xeTex) [process, right of=cleanedTexFiles, xshift=0.7cm, yshift=-0.5cm] {\xetex};
    \node (pdfTex) [process, above of=xeTex, yshift=0.4cm] {\pdftex};
    \node (luaTex) [process, below of=xeTex, yshift=0.2cm] {\luatex};

    \node (xeTexPdf) [file, right of=xeTex, xshift=0.2cm] {PDF};
    \node (pdfTexPdf) [file, right of=pdfTex, xshift=0.2cm] {PDF};
    \node (luaTexPdf) [file, right of=luaTex, xshift=0.2cm] {PDF};

    \node (xeTexImg) [file, right of=xeTexPdf, xshift=-0.1cm, yshift=0.1cm] {JPEG};
    \node (pdfTexImg) [file, right of=pdfTexPdf, xshift=-0.1cm, yshift=-0.1cm] {JPEG};

    \node (luaTexDots) [file, right of=luaTexPdf, xshift=-0.1cm] {...};

    \node (xeTexText) [file, right of=xeTexImg, xshift=-0.2cm, yshift=-0.5cm] {Text};
    \node (pdfTexText) [file, right of=pdfTexImg, xshift=-0.2cm, yshift=0.5cm] {Text};

    \node (cmpPixel) [process, above of=xeTexPdf, yshift=-0.4cm] {Pixel};
    \node (cmpFeatures) [process, above of=xeTexImg, yshift=-0.5cm] {SIFT};
    \node (cmpText) [process, above of=xeTexText, xshift=0.6cm, yshift=-0.1cm] {Levenshtein};
    \node (cmpFont) [process, above of=xeTexText, xshift=-0.3cm, yshift=0.3cm] {Font};

    \node (analysis) [misc, right of=arxivSource, xshift=2cm, yshift=0.2cm] {Root cause analysis};

    % lines ------------------------------------
    \draw[arrow] (arxivSource) -- (compressedArxivSrc) node[midway, above] {/GET};
    \begin{scope}[transform canvas={xshift=-0.6cm}]
      \draw[arrow] (compressedArxivSrc) -- (extractedArxivSrc) node[midway, right] {Extract};
      \draw[arrow] (extractedArxivSrc) -- (cleanedTexFiles) node[midway, right, text width=1.2cm] {Recondition};
      \draw[arrow] (cleanedTexFiles) -- (texEntrypoint) node[midway, right, text width=1.2cm] {Identify};
    \end{scope}

    \draw[arrow] (texEntrypoint.east) -- (xeTex.west);
    \draw[arrow] (texEntrypoint.east) -- (pdfTex.west);
    \draw[arrow] (texEntrypoint.east) -- (luaTex.west);

    \draw[arrow] (xeTex.east) -- (xeTexPdf.west);
    \draw[arrow] (pdfTex.east) -- (pdfTexPdf.west);
    \draw[arrow] (luaTex.east) -- (luaTexPdf.west);

    \draw[arrow] (pdfTexPdf) -- (pdfTexImg);
    \draw[arrow] (pdfTexPdf) -- (pdfTexText);
    \draw[arrow] (xeTexPdf) -- (xeTexImg);
    \draw[arrow] (xeTexPdf) -- (xeTexText);
    \draw[arrow] (luaTexPdf) -- (luaTexDots);

    \draw[arrow] (xeTexPdf) -- (cmpPixel);
    \draw[arrow] (pdfTexPdf) -- (cmpPixel);
    \draw[arrow] (xeTexImg) -- (cmpFeatures);
    \draw[arrow] (pdfTexImg) -- (cmpFeatures);
    \draw[arrow] (xeTexText) -- (cmpFont);
    \draw[arrow] (pdfTexText) -- (cmpFont);
    \draw[arrow] (xeTexText) -- (cmpText);
    \draw[arrow] (pdfTexText) -- (cmpText);

    \draw[arrow] ($(pdfTexPdf.north)+(0,0.6)$) |- ($(analysis.west)+(-0.4,0)$) |- (analysis.west);
    \draw[arrow] ($(luaTexDots.east)+(1.8,0)$) |- ($(luaTexDots.east)+(2.5,0)$) |- ($(analysis.east)+(0.3,0)$) |- (analysis.east);

    % rectangles ---------------------------------
    \draw[red,thick,dotted] ($(compressedArxivSrc.north west)+(-0.2,0.3)$)  rectangle ($(texEntrypoint.south east)+(0.2,-0.3)$) node[at start, right, yshift=-0.15cm] {Processing};
    \draw[red,thick,dotted] ($(pdfTex.north west)+(-0.4,0.7)$)  rectangle ($(luaTex.south east)+(0.3,-0.2)$) node[at start,right, yshift=-0.3cm, text width=1.5cm] {Compilation (\texttt{latexmk})};
    \draw[red,thick,dotted] ($(pdfTexPdf.north west)+(-0.1,0.6)$)  rectangle ($(xeTexText.south east)+(1.2,-0.1)$) node[at start,right, yshift=-0.3cm, text width=1.5cm] {Comparison (\pdftex/\xetex)};
    \draw[red,thick,dotted] ($(xeTexPdf.north west)+(-0.2,0.2)$)  rectangle ($(luaTexDots.south east)+(1.8,-0.1)$) node[at end,left, yshift=0.3cm, text width=1.5cm] {Comparison (\xetex/\luatex)};

  \end{tikzpicture}
  \caption{Overview of the automated pipeline}
  \label{fig:approachOverview}
\end{figure}

We obtained \tex sources by retrieving papers with downloadable source code from \url{arxiv.org}.
As detailed in the rest of this section, we
(1) extracted the compressed sources; (2) cleaned the \tex files by removing patterns that result in expected cases of cross-engine incompatibilities; and (3) identified entry points to compile the \tex sources from.
The PDFs compiled by \pdftex, \xetex, and \luatex were then compared to detect differences across engines. 

\paragraph{Reconditioning} 

Each \tex engine provides engine-specific primitives which may cause compilation errors with other \tex engines. 
Files that use these primitives are thus not engine-agnostic and are expected to produce different output on different \tex engines.
Instead of discarding these files, we perform a reconditioning step (as described in~\citet{Lecoeur:ReconditioningUndefinedBehaviour}) to avoid undefined behaviour and expected compilation failures.
By ``cleaning'' the \tex sources to remove engine-specific primitives such as the \texttt{\textbackslash pdffilesize} primitive that is supported by \pdftex but not \xetex or \luatex, we aim to reduce the rate of false positives while still maintaining a high volume and diversity of input files.

\paragraph{Compilation across \tex engines}

To answer \ref{rq1:interengine}, we compiled each file with \pdftex, \xetex, and \luatex in the 2023 distribution of \texlive.
The \texttt{latexmk} program was used to detect and perform the necessary compilation steps instead of invoking the \tex engines directly as some files may require multiple runs or different commands (for instance, due to the use of Bib\TeX).

\paragraph{Compilation across \texlive distributions} 

To answer \ref{rq2:interversion}, we compiled each file in the 2020, 2021, 2022, and 2023 release of \texlive using a minimal \texttt{latexmk} configuration.
The different distributions of \texlive were obtained from the official Docker images of historic \texlive releases~\cite{TeXLiveDocker}.

\paragraph{Comparison}

Pairwise comparisons were performed between PDF files generated from the same input \tex source. 
For \ref{rq1:interengine}, \xetex was chosen as the basis of comparison (\emph{i.e.}, we compared \xetex against \pdftex, and \xetex against \luatex) because our initial investigation showed that the diffs produced from the \pdftex/\luatex comparison were typically similar to the diffs produced from the \pdftex/\xetex comparison.
This was because both \xetex and \luatex are Unicode-based, resulting in similar handling of fonts.

For \ref{rq2:interversion}, we performed pairwise comparisons between adjacent versions of \texlive (2020 versus 2021, 2021 versus 2022, 2022 versus 2023).
Additionally, we compared the earliest and latest of these four releases (\emph{i.e.}, 2020 versus 2023) to understand how documents changed over a longer period of time.
A sample of inconsistencies was selected to be analysed in \ref{rq3:bugs}, detailed in Section~\ref{sec:methodology:rootCauses}.

\subsection{Characterising Differences}\label{sec:methodology:characterisingDifferences}

Building on the findings from our initial study, we refined the four methods of comparison to identify the different types of inconsistencies across engines, detailed in the rest of this section.

\paragraph{Pixel-wise comparison} 

We performed page-by-page pixel-wise comparisons using the \texttt{diff-pdf} CLI tool.
This was aimed at detecting small changes in character spacing or other inconsistencies that would be otherwise hard to notice by the human eye.

\paragraph{Text comparison} 

From our initial study, we found that pixel-wise comparisons had limited effectiveness in identifying the kinds of differences found, as the inconsistencies found in pixel-wise comparisons mostly comprised of differences in line breaks and hyphenation, as well as minor differences in text rendering.
We used a text-based comparison method as a more fine-grained tool to detect a wider variety of inconsistencies and more subtle differences in text content, as explained in the motivating study.

We first extracted the text from each PDF using the \texttt{PyMuPDF} library.
To overcome the prevalence of false positives in the motivating study, the extracted text was further processed to ignore differences that would be transparent to users, such as differences in how whitespace is encoded.
We then performed pairwise comparisons between processed texts using the Levenshtein edit distance and unique characters present.
A non-zero edit distance was used to detect the presence of inconsistencies. Comparing the unique characters present in each PDF, as well as the difference in the number of characters, helped to identify the type of inconsistencies.

\paragraph{Font-based comparison} 

To detect text formatting inconsistencies, we extracted the font information from the text in each PDF using the \texttt{PyMuPDF} library.
Due to the different font formats and encodings, the exact fonts used differ between engines and we cannot directly compare the fonts used.
To overcome this, we also compared the number of fonts present as a proxy for the number of unique styles present in the document, along with font sizes and colours.

\paragraph{Feature-based comparison}

Since text-based comparison methods cannot identify non-text inconsistencies and a pixel-wise comparison yielded a high rate of false positives in our initial study, we used a feature-based comparison to identify additional inconsistencies.

For each PDF, the first 3 and last 3 pages were converted to JPEG images, and comparison was done using the SIFT algorithm in \texttt{opencv}.
This subset of pages was chosen to reduce the computational load of the feature extraction step as we found that there was diminishing marginal utility in comparing a larger number of pages, especially for very long documents (\emph{e.g.}, more than 20 pages) where differences in text spacing made page-wise comparisons no longer meaningful.
We selected the first and last pages of each document as we observed that these often had the most variation in formatting, 
with title pages, tables of content, and section titles in the first few pages,
as well as references and appendices (including tables and figures) in the final pages.

For each pairwise comparison, a similarity score (from 0 to 1) was calculated to represent how similar the two images are.
A score of 0.7 was empirically chosen to be the threshold of considering two images to be significantly inconsistent.

\subsection{Identifying Root Causes and Triage}\label{sec:methodology:rootCauses}

For each document identified to have inconsistencies, we manually categorised the types of inconsistency observed.
We further analysed the root causes of disproportionately high occurrences of any type of inconsistency.
By doing so, we aim to identify common inconsistencies that authors are likely to come across, instead of inconsistencies that could be due to authors' idiosyncrasies.

To answer \ref{rq3:bugs}, we studied the root causes of inconsistencies in a sample of documents.
We selected all documents with either compilation failures, inconsistencies introduced only in 2023, or inconsistencies reverted only in 2023.
Doing so helped to achieve a diverse sample of inconsistencies, including compile failures, that remain relevant in the current distribution of \texlive.
This accounted for a total of 26 documents (comprising 7 compilation failures, 13 with new inconsistencies, and 6 with reverted inconsistencies).
We manually reduced the test case~\cite{Zeller:deltaDebugging}, and inspected log files and program outputs, to isolate the cause of inconsistencies; using this approach, we identified unique root causes by verifying the minimal change required to reproduce and eliminate each inconsistency.
Finally, we filed bug reports for unexpected behaviours.

\paragraph{Alignment study}

Since manual classification and analysis is potentially subjective, we used \emph{``negotiated agreement''}~\cite{Campbell:Sociological, Rabe:trivialPackages} to agree on benchmarks and arrive at a consensus.

For identifying and characterising inconsistencies (\ref{rq1:interengine}, \ref{rq2:interversion}), we reviewed the 10 \tex sources from our motivating study, yielding 20 pairwise comparisons.
Both authors independently listed and categorised the inconsistencies observed.
Using the ``negotiated agreement'' technique, both authors discussed all divergent classifications and reached a consensus for categorising the rest of the data.
Given the lack of existing studies on comparing inconsistent PDFs, this approach seemed reasonable.

A similar approach was taken to mitigate error in identifying the root causes of differences.
For all 26 cases studied in \ref{rq3:bugs}, both authors analysed and agreed on the root cause of inconsistencies.

%% file: sections/results-1.tex
\section{Results}

We found significant inconsistencies in the outputs of different \tex engines and different \texlive distributions: only 0.2\% of documents compiled to identical output with \xelatex and \pdflatex, while only 42.1\% of produced the same output from \texlive 2020 to 2023.
From further analysing a sample of documents, we identified two new bugs in \latex packages, and five existing bugs that were fixed independently of this study.

\subsection{\ref{rq1:interengine}: \tex Engines Differences}

The results obtained for each \tex engine differed significantly.
The highest rate of successful compilation was observed with \pdftex, and it was not uncommon for a paper to successfully compile with \pdftex but fail to compile on \xetex or \luatex (see Figure~\ref{fig:texEngineCompileResults}).
79.2\% of documents compiled on all three engines, and 19.4\% failed to compile only with \xetex.

\begin{figure}[tb]
  \begin{minipage}[c]{0.57\columnwidth}
    \centering
    \setlength{\tabcolsep}{2pt}%
    \captionof{table}{Compilation success rates}
    \begin{tabular}{lrrr}\toprule 
        & \multicolumn{3}{c}{\tex engine} \\ \cmidrule(lr){2-4}
      Result/\%  & \pdftex & \xetex & \luatex  \\\midrule
      Success            & 99.1    & 79.2   & 98.6  \\
      Failure            & 0.9     & 20.8   & 1.4   \\\bottomrule
    \end{tabular}%
    \label{fig:texEngineCompileResults}
  \end{minipage}
  \begin{minipage}[c]{0.39\columnwidth}
    \centering
    \setlength{\fboxsep}{0pt}
    \fbox{ \begin{venndiagram3sets}[
        labelA={\pdftex\;}, labelB={\;\xetex}, labelC=\luatex,
        labelOnlyB={0}, labelOnlyC={0}, labelOnlyAB={0}, labelOnlyBC={0},
        labelABC={342}, labelOnlyAC={84}, labelOnlyA={2}, 
        % labelNotABC={\quad 4},
        vgap=0cm, hgap=0cm, radius=0.8cm, overlap=0.8cm, showframe=false
        ]%
        % HACK: instead of labelNotABC={4}, this saves a bit of vertical space
        \node(NotABC)[inner sep=0pt, text width=0.2cm, text height=0]{\small 4};%
    \end{venndiagram3sets} } %
    \caption{Number of successful compilations}
    \label{fig:texEngineCompileResultsVenn}
  \end{minipage}
\end{figure}

% \begin{table}[tb]\centering
%   \caption{Compilation results (\%) for each \tex engine}
%   \label{tbl:texEngineCompileResults}
%   \begin{tabular}{lrrr}\toprule 
%   & \multicolumn{3}{c}{\tex engine} \\ \cmidrule(lr){2-4}
%     Compile result/\%  & \pdflatex & \xelatex    & \lualatex  \\\midrule
%     PDF produced with no errors    & 72.0 & 63.0  & 59.5  \\
%     PDF produced with errors       & 27.1 & 16.2  & 39.1  \\
%     No PDF produced                & 0.9  & 20.8  & 1.4   \\\bottomrule
%   \end{tabular}
% \end{table}

Compilation failures generally occurred due to incompatibilities between \latex packages and \tex engines, not due to \tex engine bugs.
We identified the following root causes for compilation failures:
\begin{enumerate*}
  \item a \latex package used is incompatible with some \tex engine, or requires action from the author to ensure compatibility (such as including a driver);
  \item the \latex document class is incompatible with a \tex engine;
  \item a syntax is present that relies on a specific \tex engine;
  \item and the source file has a syntax error.
\end{enumerate*}

\begin{figure}[tb]
  \Description[Different text spacing in PDFTeX and XeTeX]{Example of PDFTeX and XeTeX compiling the same input document with different text spacing}
  \setlength{\fboxsep}{0pt}%
  \begin{subfigure}{.48\columnwidth}
    \centering
    \fbox{\includegraphics[width=0.87\linewidth]{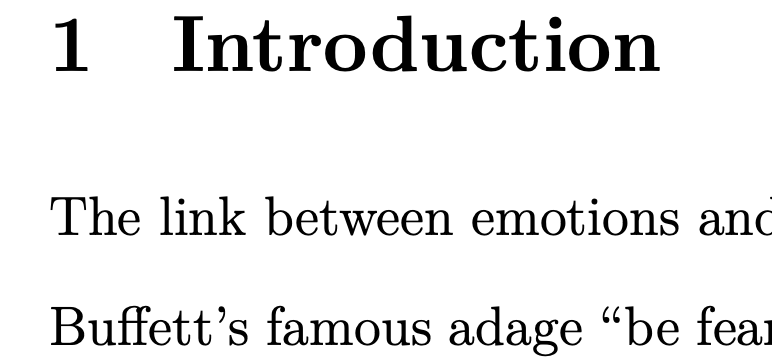}}
    \caption{PDF generated by \xelatex}
    \label{fig:sampleDiff:textSpacing:xetex}
  \end{subfigure}
  \begin{subfigure}{.48\columnwidth}
    \centering
    \fbox{\includegraphics[width=0.87\linewidth]{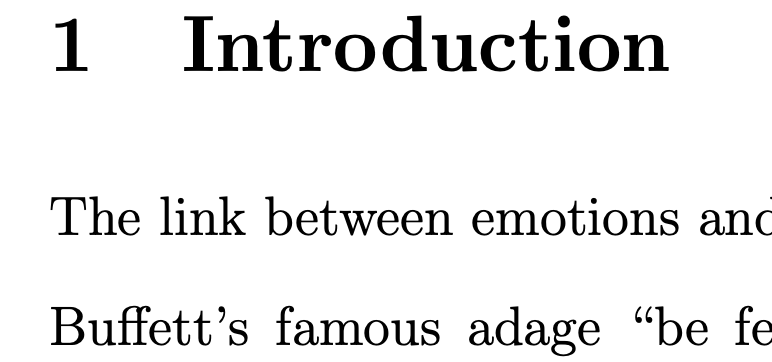}}
    \caption{PDF generated by \pdflatex}
    \label{fig:sampleDiff:textSpacing:pdftex}
  \end{subfigure}

  \begin{subfigure}{0.98\columnwidth}
    \centering
    \fbox{\includegraphics[width=0.45\linewidth]{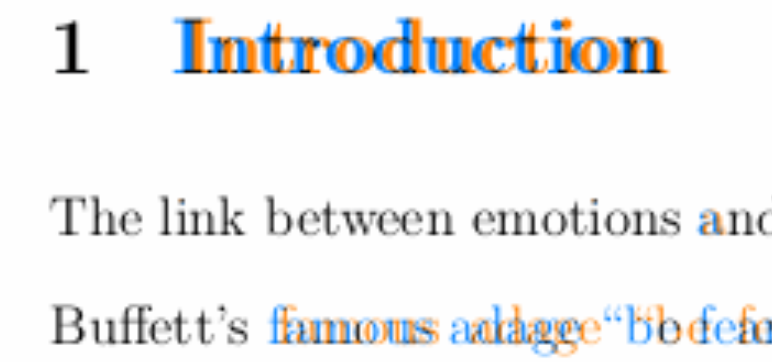}}
    \caption{Overlaying PDFs to highlight differences
      (\colorbox{cyan}{\textcolor{cyan}{X}} \xetex; \ 
      \colorbox{orange}{\textcolor{orange}{X}} \pdftex)
    }
    \label{fig:sampleDiff:textSpacing:diffXePdf}
  \end{subfigure}

  \caption{Different text spacing (arXiv:2306.12602~\cite{arxiv:2306.12602})}
  \label{fig:sampleDiff:textSpacing}
\end{figure}

\begin{figure}[tb]
  \Description[Different ligatures in PDFTeX and XeTeX]{Example of PDFTeX and XeTeX compiling the same input document with different ligatures for quotation marks}
  \setlength{\fboxsep}{0pt}%
  \begin{subfigure}{.48\columnwidth}
    \centering
    \fbox{\includegraphics[width=0.90\linewidth]{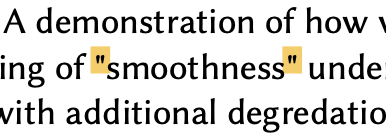}}
    \caption{PDF generated by \pdflatex}
    \label{fig:sampleDiff:ligatures:pdftex}
  \end{subfigure}
  \begin{subfigure}{.48\columnwidth}
    \centering
    \fbox{\includegraphics[width=0.90\linewidth]{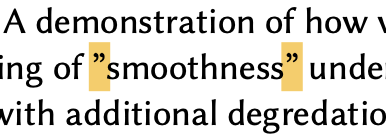}}
    \caption{PDF generated by \xelatex}
    \label{fig:sampleDiff:ligatures:xetex}
  \end{subfigure}

  \caption{Different quotation marks (arXiv:2306.01691~\cite{arxiv:2306.01691})}
  \label{fig:sampleDiff:ligatures}
\end{figure}

\begin{figure}[tb]
  \Description[Different references in LuaTeX and XeTeX]{Example of LuaTeX and XeTeX compiling the same input document with different references}
  \setlength{\fboxsep}{0pt}%
  \begin{subfigure}{.48\columnwidth}
    \centering
    \fbox{\includegraphics[width=0.98\linewidth]{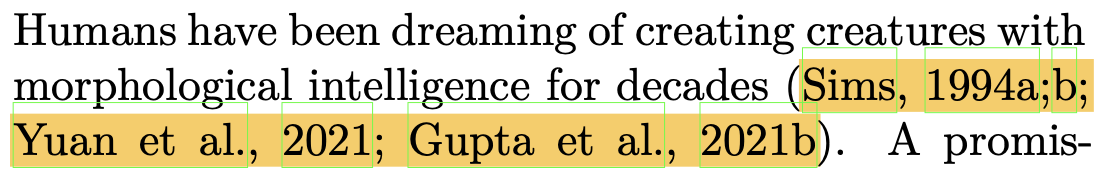}}
    \caption{PDF generated by \xelatex}
    \label{fig:sampleDiff:references:xetex}
  \end{subfigure}
  \begin{subfigure}{.48\columnwidth}
    \centering
    \fbox{\includegraphics[width=0.98\linewidth]{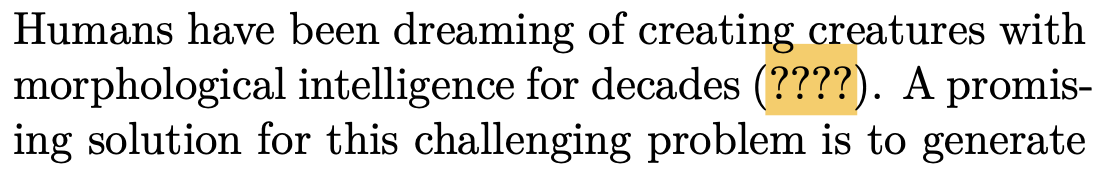}}
    \caption{PDF generated by \lualatex}
    \label{fig:sampleDiff:references:luatex}
  \end{subfigure}

  \caption{Differences in references (arXiv:2306.00036~\cite{arxiv:2306.00036})}
  \label{fig:sampleDiff:references}
\end{figure}

From the differences in successful compilation results (\emph{i.e.}, ``Different output'' in Table~\ref{tbl:pairwiseComparisonResults}), we identified several incompatibilities between \tex document classes and \tex engines.
Generally, most incompatibilities occurred with \xetex and \luatex (\emph{i.e.}, many packages were compatible only with \pdftex).
The common types of differences caused by different document classes were:

\begin{itemize}
  \item \textbf{Text spacing}, referring to the spaces or misalignments between individual characters (see Figure~\ref{fig:sampleDiff:textSpacing});
  \item \textbf{Line breaks}, referring to how text (\emph{e.g.}, a paragraph) is divided into multiple lines (see Figure~\ref{fig:sampleDiff:wordBreaks});
  \item \textbf{Number of pages} in the output PDF;
  \item \textbf{Missing content}, where one output PDF contains content that another does not (see Figure~\ref{fig:sampleDiff:headerContent});
  \item \textbf{Inconsistent styles}, such as italics or bold text (Figure~\ref{fig:sampleDiff:missingStyles01}), differences in ligatures, or differences in how a character is rendered (Figure~\ref{fig:sampleDiff:ligatures});
  \item \textbf{References}, where citations or bibliographies are rendered differently (see Figure~\ref{fig:sampleDiff:references}); and
  \item \textbf{Images}, such as the size and position of images.
\end{itemize}

\begin{table}[tb]\centering
  \caption{Pairwise comparison results (\%)}
  \label{tbl:pairwiseComparisonResults}
  \begin{tabular}{lrr}\toprule 
  & \multicolumn{2}{c}{Comparing \xelatex with:} \\ \cmidrule(lr){2-3}
    Result/\%  & \pdflatex & \lualatex  \\\midrule
    Compile failure    & 20.8  & 20.8   \\
    Different output   & 78.9  & 77.8   \\
    Consistent output  & 0.2   & 1.4    \\\bottomrule
  \end{tabular}
\end{table}

\begin{table*}[tb]
  \caption{Differences across common document classes (\xelatex vs \pdflatex)}
  \label{table:diffResultsByDocumentclass:XePdf}
  \begin{tabular}{lrrrrrrr}\toprule 
    & \multicolumn{7}{c}{\% of papers with difference (\xelatex versus \pdflatex)} \\ \cmidrule(lr){2-8}
    Class (count)      & Missing styles & Missing content & Number of pages & Images & Text spacing & Line breaks & References  \\\midrule
    All compiled (342)  & 21.9  & 2.3   & 17.8  & 27.8  & 98.3  & 96.2  & 0.9  \\\midrule
    {amsart}     (65)   & 6.2   & 0.0   & 21.5  & 21.5  & 96.8  & 88.9  & 0.0  \\
    {revtex4-2}  (43)   & 7.0   & 0.0   & 25.6  & 37.2  & 95.3  & 95.3  & 2.3  \\
    {IEEEtran}   (27)   & 63.0  & 0.0   & 63.0  & 70.4  & 100.0 & 88.9  & 0.0  \\
    {elsarticle} (26)   & 19.2  & 0.0   & 30.8  & 50.0  & 100.0 & 100.0 & 0.0  \\
    {revtex4-1}  (23)   & 0.0   & 0.0   & 30.4  & 52.2  & 100.0 & 100.0 & 0.0  \\
    {acmart}     (21)   & 19.0  & 0.0   & 28.6  & 52.4  & 95.0  & 95.0  & 0.0  \\\bottomrule
  \end{tabular}
\end{table*}

\begin{table*}[tb]
  \caption{Differences across common document classes (\xelatex vs \lualatex)}
  \label{table:diffResultsByDocumentclass:XeLua}
  \begin{tabular}{lrrrrrrr}\toprule 
    & \multicolumn{7}{c}{\% of papers with difference (\xelatex versus \lualatex)} \\ \cmidrule(lr){2-8}
    Class (count)      & Missing styles & Missing content & Number of pages & Images & Text spacing & Line breaks & References  \\\midrule
    All compiled (340)  & 0.0   & 4.1   & 8.2   & 19.1  & 74.8  & 74.3  & 10.0 \\\midrule
    {amsart}     (65)   & 0.0   & 12.3  & 30.8  & 21.5  & 50.8  & 44.4  & 21.5 \\
    {revtex4-2}  (43)   & 0.0   & 0.0   & 23.3  & 37.2  & 74.4  & 76.7  & 4.7  \\
    {IEEEtran}   (27)   & 0.0   & 3.7   & 29.6  & 37.0  & 88.9  & 88.9  & 3.7  \\
    {elsarticle} (26)   & 0.0   & 0.0   & 23.1  & 38.5  & 80.0  & 80.0  & 0.0  \\
    {revtex4-1}  (23)   & 0.0   & 0.0   & 30.4  & 52.2  & 73.9  & 73.9  & 0.0  \\
    {acmart}     (21)   & 0.0   & 0.0   & 23.8  & 47.6  & 90.0  & 95.0  & 0.0  \\\bottomrule
  \end{tabular}
\end{table*}

The occurrences of these differences are summarised in Table~\ref{table:diffResultsByDocumentclass:XePdf} (\xetex against \pdftex) and Table~\ref{table:diffResultsByDocumentclass:XeLua} (\xetex against \luatex).
From analysing document classes with a disproportionately high rate of inconsistencies, 
we identified several popular document classes that are compatible with \pdftex but incompatible with other engines.
For instance, due to differences in how each engine handles font loading, the \texttt{IEEEtran} class is compatible with \pdftex only; all styles (such as italics, bold, and small-caps) are not rendered in other engines (Figure~\ref{fig:sampleDiff:missingStyles01}).
\texttt{IEEEtran} is the template recommended by IEEE (Institute of Electrical and Electronics Engineers) for use in conference proceedings~\cite{ieeetran:latexTemplate}.
Similar incompatibilities were observed, though to a lesser extent, with the \texttt{mnras} class (for the Monthly Notices of the Royal Astronomical Society~\cite{mnras:latexTemplate}).

\begin{figure*}[tb]
  \centering
  \begin{subfigure}{0.48\linewidth}
    \centering
    \Description{Venn diagram of the distribution of inconsistencies observed}
    % \centering
    \includegraphics[width=0.97\linewidth]{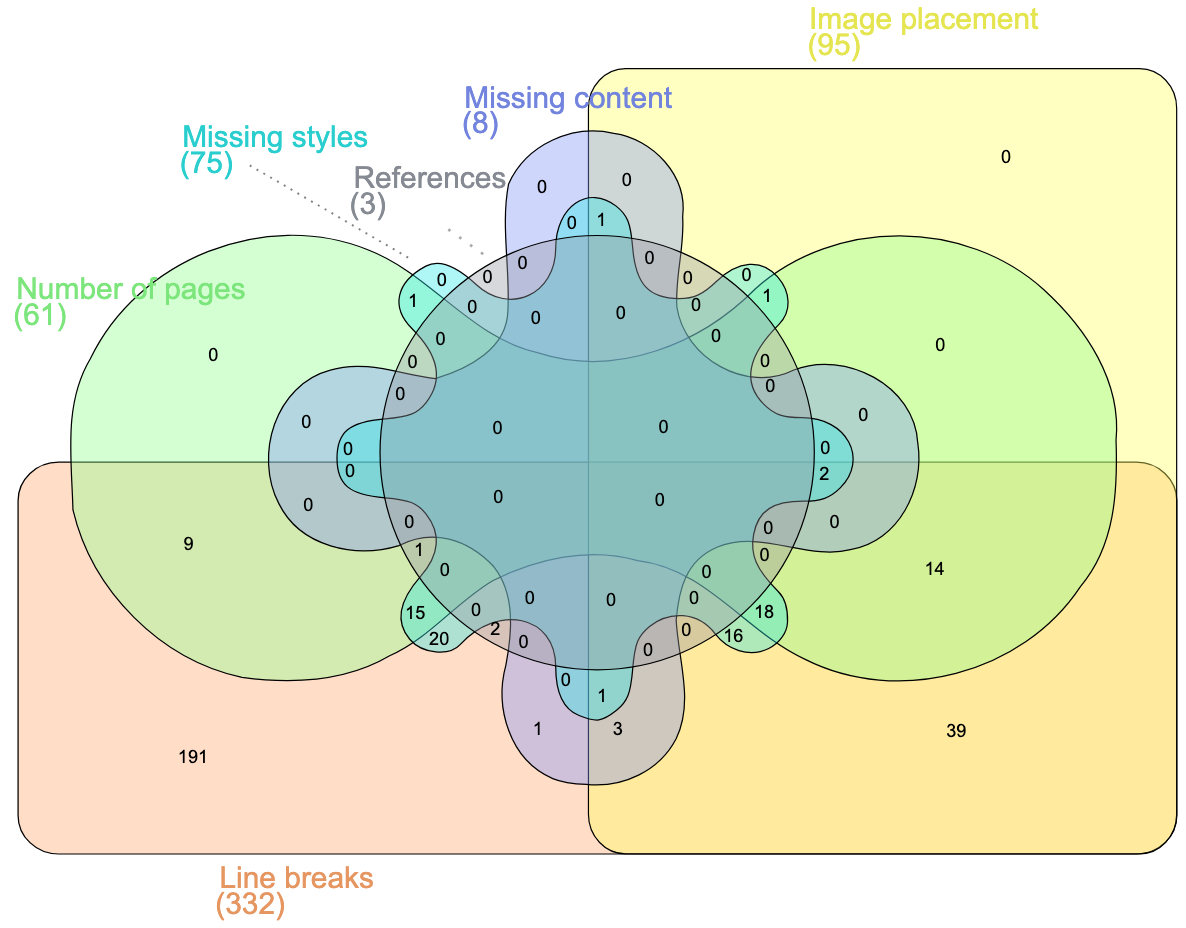} 
    \caption{Distribution of inconsistencies between \pdftex and \xetex}
    \label{fig:vennDiagram:xepdf}
  \end{subfigure}
  \begin{subfigure}{0.48\linewidth}
    \centering
    \Description{Venn diagram of the distribution of inconsistencies observed}
    % \centering
    \includegraphics[width=0.97\linewidth]{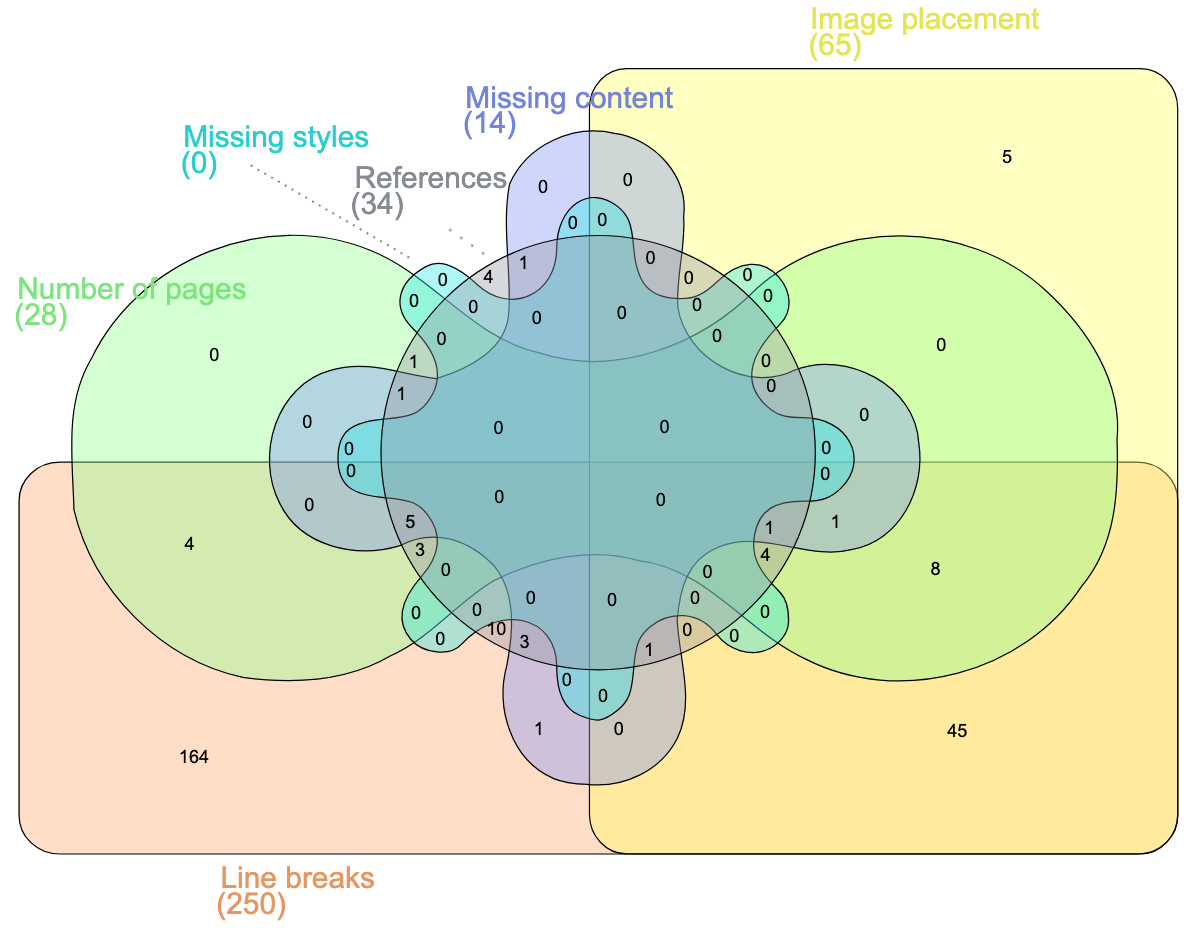} 
    \caption{Distribution of inconsistencies between \luatex and \xetex}
    \label{fig:vennDiagram:xelua}
  \end{subfigure}
  \caption{Distribution of inconsistencies in pairwise comparisons of \tex engines}
\end{figure*}

The Venn diagram in Figure~\ref{fig:vennDiagram:xepdf} illustrates how the different inconsistencies are distributed when comparing \pdftex and \xetex.
We observed that relatively higher numbers (15 or greater) of documents containing a given set of inconsistencies tended to occur more often compared to smaller numbers (3 or below).
This is likely because inconsistencies tend to occur based on the document class of the \tex file, and thus input sources using the same document class would have the same types of inconsistencies.
Small numbers in the Venn diagram were likely due to documents where the author imported a specific \latex package that was incompatible with some \tex engine, or due to the author using engine-specific syntax that triggered a specific type of inconsistency not usually observed with the document class.
A similar trend was also observed when comparing \xetex and \luatex (Figure~\ref{fig:vennDiagram:xelua}).

%% file: sections/results-2.tex
\subsection{\ref{rq2:interversion}: \texlive Version Differences}\label{sec:results2}

\paragraph{Overview} 

We found significant differences in the compilation results of different versions of \texlive.
Similar to the inter-engine comparison, the most common type of inconsistencies were differences in text spacing, which were usually relatively minor:
such differences were mostly small shifts in the position of characters on a page, and did not affect the information conveyed in the document.
However, more ``major'' differences affecting the content of the document were still prevalent.

Table~\ref{tbl:versionCompareResuls} shows the pairwise comparison results across four years from 2020 to 2023 (the supplementary materials provide more detailed statistics on the inconsistencies in each year).
Notably, \texlive 2021 to 2023 was mostly stable, with over 80\% of documents compiling to the same output across all three years.
However, the shift from \texlive 2020 to \texlive 2021 resulted in significant discrepancies across most documents, where only 20.4\% of documents produced the same output across both versions of \texlive.

\paragraph{Breaking changes in compilation flags} 

From 2020 to 2021, a significant proportion (48.6\%) of documents had discrepancies in references.
This was due to a change in the default value of the \texttt{bibtex\_fudge} option used by \texttt{latexmk}, an automated script included in the \texlive distribution for compiling \latex documents.
This regression affected the compilation of documents with Bib\tex citations, and was patched in the subsequent version of \texttt{latexmk}.
After accounting for the change in compilation commands required in \texlive 2020, the proportion of documents with discrepancies in references fell from 48.6\% to only 2.8\%. 

\begin{table}[tb]\centering
  \aboverulesep=0mm \belowrulesep=0mm
  \caption{Comparison results (\%) from \texlive 2020 to 2023}
  \label{tbl:versionCompareResuls}
  \begin{tabular}{lrrr|r}\toprule 
  & \multicolumn{4}{c}{\texlive versions compared} \\ \cmidrule(lr){2-5}
    Comparison result /\%  & '20/'21 & '21/'22 & '22/'23 & '20/'23  \\\midrule
    Identical PDFs  & 20.4 & 80.1  & 85.2  & 30.3  \\
    Different PDFs  & 69.4 & 19.2  & 13.9  & 68.1  \\
    Compile failure & 1.2  & 0.7   & 0.9   & 1.6   \\\bottomrule
  \end{tabular}
\end{table}

\paragraph{Types of inconsistencies observed} 

After accounting for changes in compilation commands, the amount of inconsistencies found decreased across all categories but still remained significant, with only 54.5\% of documents compiling to the same output across both years.
Differences in text spacing were the most common discrepancy observed, affecting 47.2\% of documents; 
such differences were particularly common in document classes like \texttt{acmart} (76\%), \texttt{amsart} (51\%) and \texttt{mnras} (88\%),
as well as the packages for \texttt{neurips} (90\%) and \texttt{icml2023} (86\%).
Differences in text spacing throughout a document may result in more major inconsistencies: of the eight documents using the \texttt{mnras} class, one document compiled to a different number of pages in 2020 and 2021 due to accumulated differences in text spacing despite having no difference in text content.

Similar to the cross-engine comparison, other discrepancies were observed as well, ranging from differences in font styles to missing content and missing references.
Table~\ref{tbl:diffResultsByVersion} shows the distribution of the discrepancies observed across the years.
Notably, from 2021 to 2022, we observed a relatively higher proportion of inconsistencies related to references and text content.

\begin{table*}[tb]
  \caption{Differences across different \texlive versions (2020--2023)}
  \label{tbl:diffResultsByVersion}
  \begin{tabular}{lrrrrrrr}\toprule 
    & \multicolumn{7}{c}{\% of papers with difference (comparing \texlive versions)} \\ \cmidrule(lr){2-8}
    Versions & Missing styles & Missing content & Number of pages & Images & Text spacing & Line breaks & References  \\\midrule
    '20/'23  & 6.0  & 39.4  & 34.0  & 0.7  & 65.7  & 44.7  & 44.4 \\
    '20/'23 ($\ast$) & 5.8  & 3.5   & 7.9   & 1.6  & 51.4  & 14.8  & 11.6  \\\midrule
    '20/'21  & 5.6  & 45.4  & 38.4  & 0.0  & 67.1  & 49.8  & 48.6 \\
    '20/'21 ($\ast$) & 4.6  & 2.3   & 2.8   & 0.7  & 42.4  & 7.6   & 2.8  \\
    '21/'22  & 1.4  & 6.5   & 5.3   & 0.0  & 15.5  & 8.8   & 10.0 \\
    '22/'23  & 1.6  & 0.7   & 0.5   & 0.7  & 13.0  & 3.2   & 2.3  \\\bottomrule
  \end{tabular}

  {\small ($\ast$) Compiled with extra compilation flags due to changes in \texttt{latexmk} default options }
\end{table*}

\paragraph{Missing references} 

We observed a relatively higher incidence of inconsistencies with references from 2021 to 2022 (affecting 10.0\% of documents, compared to under 2.8\% from 2020 to 2021 and 2.3\% from 2022 to 2023).
This was largely due to a difference in \texttt{latexmk}'s automated compilation in response to syntax errors related to bibliographies.
In the 2021 \texlive distribution, \texttt{latexmk} would often continue to run compilation steps after encountering a Bib\tex error, allowing some references to be rendered in spite of syntax errors;
whereas in the 2022 distribution, \texttt{latexmk} would often stop attempting compilation, resulting in differences in the references across both years.
This change in behaviour was likely not a bug as it resulted from authors' syntax errors being handled differently, and errors were appropriately logged to warn authors.

\paragraph{Changes in output over four years} 

Table~\ref{tbl:versionComparePatternsOverTime} shows the changes in compilation output over time, with ``\yes'' denoting identical PDF output between two years and ``\no'' denoting different output.
Even after accounting for differences in compilation flags, only 42.1\% of documents had consistent output across all four versions of \texlive. 
Of the documents that did not compile to the same output in all four years, the majority had differences in output only between the 2020 and 2021 versions (accounting for 27.1\% of the total).
The relatively high rate of inconsistencies from 2020 to 2021 is largely attributed to differences in text spacing as described above.
Interestingly, Table~\ref{tbl:versionComparePatternsOverTime} shows how some changes were reverted in later distributions of \texlive, with 1.7\% of documents having consistent output when compiled with the 2020 and 2023 versions, but not intermediate versions.
We further discuss these cases in Section~\ref{sec:results3}.

\begin{table}[tb]\centering
  \caption{Comparison results over time (2020 to 2023)}
  \label{tbl:versionComparePatternsOverTime}
  \aboverulesep=0mm \belowrulesep=0mm
  \yes\, identical output; \; \no\, different output
  \begin{tabular}{ccc|c|r}\toprule 
    \multicolumn{4}{c|}{Comparison results} & \multirow{2}{*}{Count} \\ \cmidrule{1-4}
    '20/'21 & '21/'22 & '22/'23 & '20/'23   \\\midrule
    \yes	& \yes	& \yes	& \yes	& 42.1\% \\
    \no 	& \yes	& \yes	& \no 	& 27.1\% \\
    \no 	& \no 	& \yes	& \no 	& 8.1\%  \\
    \yes	& \no 	& \yes	& \no 	& 7.9\%  \\
    \no 	& \yes	& \no 	& \no 	& 6.7\%  \\
    \yes	& \yes	& \no 	& \no 	& 3.0\%  \\
    \no 	& \no 	& \no 	& \no 	& 2.8\%  \\
    \no 	& \yes	& \no 	& \yes	& 1.2\%  \\
    \yes	& \no 	& \no 	& \no 	& 0.7\%  \\
    \yes	& \no 	& \no 	& \yes	& 0.5\%  \\\bottomrule
  \end{tabular}
\end{table}

%% file: sections/results-3.tex
\subsection{\ref{rq3:bugs}: Root Cause Analysis} \label{sec:results3}

Overall, we analysed 26 documents (the specific documents selected are listed in the supplementary materials) and identified 10 unique root causes for the inconsistencies found.
Of the 10 root causes, two were new bugs identified by this study; five were bugs fixed independently of this study; three were expected behaviour. 

\paragraph{New bugs} 

\begin{figure}[tb]
  \Description[Different position of math equations]{Example of TeX Live 2020 and 2021 compiling the same input document with different positions of math equations}
  \setlength{\fboxsep}{0pt}%
  \begin{subfigure}{.8\columnwidth}
    \centering
    \fbox{\includegraphics[width=0.9\linewidth]{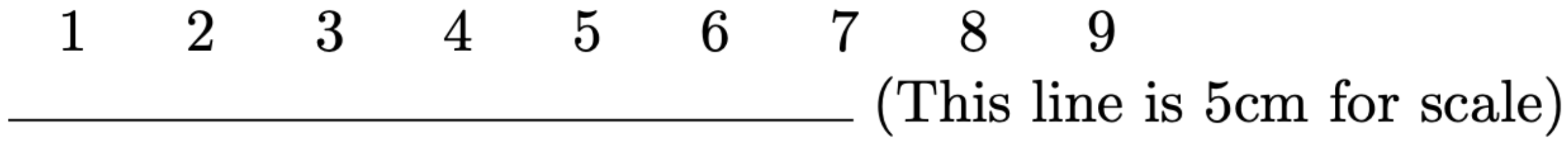}}
    \caption{Original behaviour (without importing \texttt{colortbl})}
    \label{fig:colortbl:noimport}
  \end{subfigure}
  \begin{subfigure}{.8\columnwidth}
    \centering
    \fbox{\includegraphics[width=0.9\linewidth]{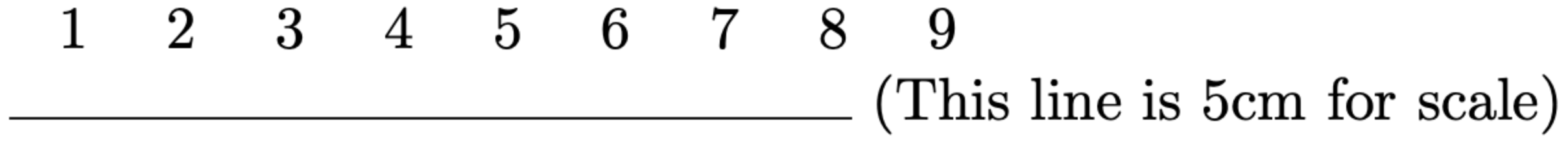}}
    \caption{Importing \texttt{colortbl}}
    \label{fig:colortbl:import}
  \end{subfigure}

  \caption{Decreased spacing when \texttt{colortbl} is imported}
  \label{fig:colortbl}
\end{figure}

In first bug we identified, we found that importing the \texttt{colortbl} package decreased the horizontal spacing within tables (shown in Figure~\ref{fig:colortbl}), even without using any commands or functionality provided by the package.
The developers responded to our bug report and confirmed that this behaviour is a bug due to inconsistent behaviour of the core \latex \texttt{array} package~\cite{colortbl:issue48, LaTeX2e:issue1323}.

The second bug was with the \texttt{jmlr} document class:
using this class resulted in a compilation failure in 2023 as it had not been updated to work with the \texlive 2023 distribution.
We filed a bug report that has yet to be triaged~\cite{jmlr:bugReport}.

\paragraph{Existing bugs} 

Our analysis identified five existing (\emph{i.e.}, identified or fixed independently of this study) bugs.
The first bug was in the core \latex package, where inserting newlines in footnotes using the ``\texttt{\textbackslash\textbackslash}'' macro would fail compilation despite having valid syntax.
This bug was reported and fixed in \texlive 2021~\cite{LaTeX2e:issue548}.
% https://www.latex-project.org/news/latex2e-news/ltnews33.pdf, "Make \\ generally robust"

The second bug was caused by the \texttt{hyperref} package, where importing the package resulted in increased spacing between an equation environment and a theorem environment.
However, the change in spacing was desirable (\emph{i.e.}, the behaviour in 2023 was ``more correct''), as it was caused by fixing an existing bug rather than introducing a new bug.

\begin{figure}[tb]
  \Description[Different position of math equations]{Example of TeX Live 2020 and 2021 compiling the same input document with different positions of math equations}
  \setlength{\fboxsep}{0pt}%
  \begin{subfigure}{.49\columnwidth}
    \centering
    \fbox{\includegraphics[width=0.99\linewidth]{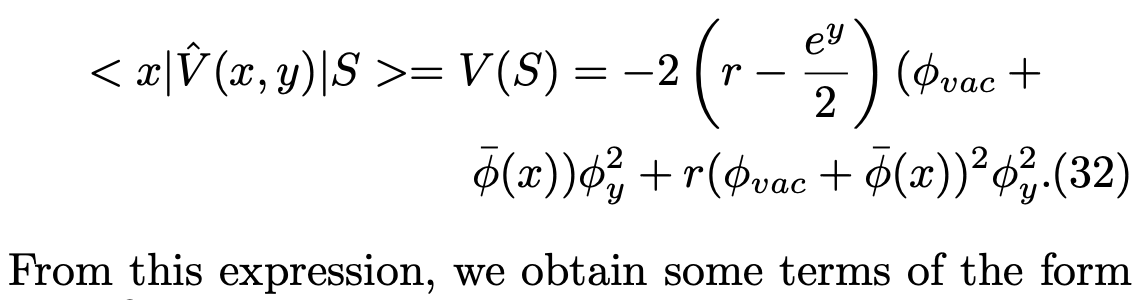}}
    \caption{Output from \texlive 2020}
    \label{fig:sampleDiff:revtexMath:2020}
  \end{subfigure}
  \begin{subfigure}{.49\columnwidth}
    \centering
    \fbox{\includegraphics[width=0.99\linewidth]{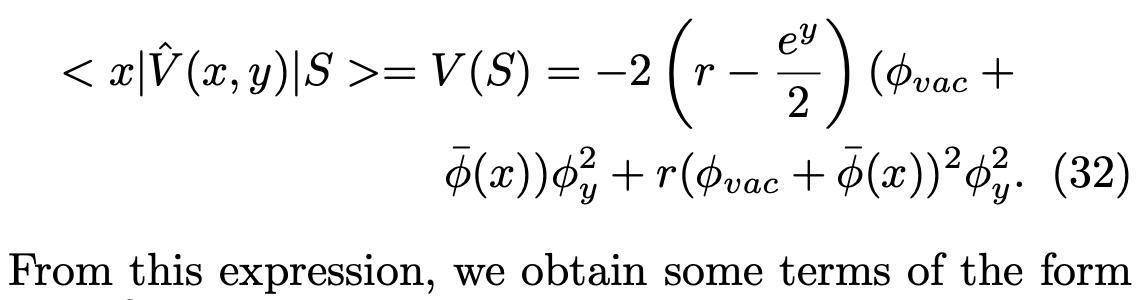}}
    \caption{Output from \texlive 2021}
    \label{fig:sampleDiff:revtexMath:2021}
  \end{subfigure}

  \caption{Math equations rendered slightly leftwards in \texlive 2021, compared to \texlive 2020 (arXiv:2306.03237~\cite{arxiv:2306.03237})}
  \label{fig:sampleDiff:revtexMath}
\end{figure}

The third bug was in the \texttt{revtex4-2} document class. We observed three documents from the \texttt{revtex4-2} document class compiled to the same output in \texlive 2020 and 2023, but \texlive 2021 compiled with slightly misaligned equations in the \texttt{eqnarray} environment (Figure~\ref{fig:sampleDiff:revtexMath}), which was the environment recommended in the REV\tex 4.2 Author's Guide~\cite{REVTexGuide}.
This change was due to a bug in the \emph{``Detection of \textbackslash{}eqnarray in Newer LaTeX''} which was fixed in the 2023 distribution~\cite{REVTeXReleases}.
The fourth bug was similar, but occurred in the core \latex package with the behaviour of the built-in \verb|\eqno| macro in \latex when handling newlines, resulting in different indentation of the paragraph after an equation for two documents.
This change was introduced in the 2022 distribution and recognised as a bug that was fixed in the 2023 distribution~\cite{LaTeX:EqnoBug}.

\begin{figure}[tb]
  \Description[Different font weight of units]{Example of TeX Live 2020 and 2021 compiling the same input document with font weights for SI units}
  \setlength{\fboxsep}{0pt}%
  \begin{subfigure}{.48\columnwidth}
    \centering
    \fbox{\includegraphics[width=0.8\linewidth]{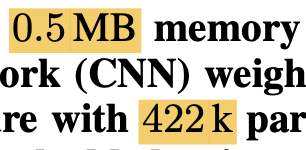}}
    \caption{Output from \texlive 2020}
    \label{fig:sampleDiff:siunitx:2020}
  \end{subfigure}
  \begin{subfigure}{.48\columnwidth}
    \centering
    \fbox{\includegraphics[width=0.8\linewidth]{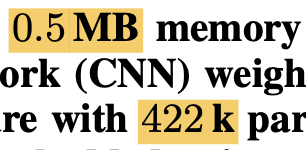}}
    \caption{Output from \texlive 2021}
    \label{fig:sampleDiff:siunitx:2021}
  \end{subfigure}

  \caption{SI units (highlighted) are bold in \texlive 2021, but not \texlive 2020 (arXiv:2306.00001~\cite{arxiv:2306.00001})}
  \label{fig:sampleDiff:siunitx}
\end{figure}

Finally, the fifth bug was caused by the \texttt{siunitx} package.
In \texlive 2021 and 2022, font styles were inconsistently applied to quantities and units as shown in Figure~\ref{fig:sampleDiff:siunitx}.
This bug was fixed in the 2023 distribution.

\paragraph{Expected behaviour} 

Three root causes of inconsistencies, accounting for 16 documents, were expected behaviour.
Three documents failed to compile in all years due to syntax errors made by authors.
Two documents failed to compile in 2020 as they imported the \texttt{tabularray} package, which was introduced only in 2021.

In the remaining 11 documents, the use of the tilde (``$\sim$'') alignment character resulted in differences in the spacing and alignment of equations from \texlive 2022 to 2023.
The developers deemed this intended behaviour as the \emph{``current behavior ... is more correct''}, and indicated that the change was due to a correction in how the tilde character was defined~\cite{LaTeX2e:issue1319}.
While the change was listed in the release notes as \emph{``Aligning status of tilde with other active characters''}~\cite{ltnews38}, the impact on authors was not directly documented.

% XXX: \ij \IJ

%% file: sections/threats-to-validity.tex
\section{Threats to Validity}

\paragraph{Dataset selection} 

Our dataset comprises only documents selected from arXiv. 
As such, there may be limited representativeness in our findings:
the distribution of inconsistencies observed may not be generalisable to other usages of \tex.
However, the types of inconsistencies observed are still applicable and may be insightful for authors wishing to understand the common inconsistencies that may occur in document preparation.

\paragraph{Bias towards \pdftex}

As seen in Figure~\ref{fig:texEngineCompileResults}, \pdftex had the highest successful compilation rate by far.
Given that \pdftex is the oldest and most popular \tex engine amongst the three, this result was unsurprising.
The high compilation rate with \pdftex may also have been because arXiv ``fully supports and automatically recognizes \pdflatex'' but not \xelatex or \lualatex~\cite{arxiv:submittingTex}, and thus the \tex sources retrieved from arXiv are more likely to be biased towards \pdflatex.

\paragraph{Identifying and characterising differences} 

When a large number of inconsistencies were present in a document, we might have missed some inconsistencies.
For instance, when a document compiled with differences in font styles, text spacing, references, and text content across different engines, it may be difficult to identify the root cause of one specific inconsistency due to the multitude of confounding factors.
We mitigated this by using a large number and variety of papers, as a diverse input sample might reveal different combinations of inconsistencies and thus allow us to better identify their root causes.
One example of this would be the \verb|\eqnarray| bug in REVTeX (see Section~\ref{sec:results2}).
This bug would have affected all 43 documents using the \texttt{revtex4-2} document class.
In three of these documents, this bug was the only inconsistency observed, allowing it to be more easily detected and diagnosed.
In the remaining 40 documents, the bug occurred alongside other inconsistencies, making it harder to both detect and characterise.
By using a large number of input files, we aim to improve the diversity of our results to obtain both isolated instances of inconsistencies, as well as cases where different inconsistencies may occur in tandem.

%% file: sections/discussion.tex
\section{Implications}

In this section, we evaluate the actionable insights of our study, and discuss how this relates to authors of \tex documents and developers in the \tex ecosystem.

\paragraph{Restricted choice of engines} 

We found that several popular document classes associated with academic journals were incompatible with \xetex and \luatex.
Thus, if an author wishes to use one of these document classes in their \tex document, they would not be able to use these \tex engines even if they took care to ensure that their document was engine-agnostic.
This is because the document class itself contained engine-specific behaviour or dependencies, and the author of a \tex document using such a class would have to modify the document class itself, which is often not allowed in contexts such as conference submissions.
This is a problem for authors who cannot use \pdflatex (for instance, \pdflatex has poor support for bidirectional text, making \xelatex or \lualatex a more popular choice for languages like Arabic~\cite{XeTeXIntro}).
Thus, while engines are purported to be \emph{``a drop-in replacement''}~\cite{TexFaq:xetexLuatex} most of the time, inconsistencies between engines are common due to the prevalence of document classes that do not support cross-engine compatibility.
Authors should thus test their documents to ensure that compilation works as intended, especially if they intend to use engines outside of \pdftex, to avoid negative outcomes (such as desk-rejection) that result from unintended output.

\paragraph{Cascading effects of incompatibilities} 

We found several incompatibilities between \latex packages and \tex engines.
Such incompatibilities have a significant impact on the \tex ecosystem as a whole: if a \latex package is incompatible with a particular \tex engine, then a document class that uses this package will likely be incompatible with the engine as well.
Such incompatibilities further limit authors' choices of \tex engines, as they may be limited to choosing \tex engines that are compatible with a given document class that they intend to use.
A reduced user base may have cascading effects on the development of a \tex engine, as exposure to a community has been shown to incentivise contributions to open-source software~\cite{Rullani2007}.

\paragraph{Effects of version upgrades} 

Over different distributions of \texlive, we observed differences in the behaviour of compilation options (such as the usage of \texttt{latexmk}), \latex packages, and document classes (including those used by popular conferences).
These discrepancies suggest that it is important for authors of \tex documents to ensure that their versions of \texlive are up-to-date, especially if they are required to adhere to a specific \latex format or document specification.
For instance, an author preparing a document using an older version of \texlive may fail to adhere to a conference's page limit when the document is compiled with the latest version, due to differences in text spacing and line breaks which result in a different number of pages.
Such differences may occur even if there have been no changes in the document class.
Authors can mitigate these risks by ensuring that their \texlive distributions are updated or aligned with the versions recommended by their submission guidelines.

\paragraph{Managing version updates} 

In spite of the importance of keeping updated to an appropriate version of \texlive, the prevalence of inter-version inconsistencies might deter authors from doing so, especially if they are unable to determine how their documents or workflows may change.
While the \latex project maintains a comprehensive changelog and release newsletter to document changes, the practical impact of changes on authors is not always directly explained.
For instance, a change in how the tilde character was internally defined was listed in the release notes as \emph{``Aligning status of tilde with other active characters''}~\cite{ltnews38}, which does not mention potential changes in how text might be rendered around the tilde character.
Due to the complexity of a typesetting system, it can be difficult to document how authors may be impacted, and changelogs may not be sufficiently informative for authors less familiar with \tex internals.
Furthermore, the impact of changes on authors is difficult to predict as changes in ``core'' functionality, such as the \latex package itself, will inevitably affect the downstream packages depending on it.
Due to the complexity of the \tex ecosystem, it is virtually impossible for authors to predict how their documents may be affected by version changes.
Authors may mitigate the risk of unexpected or undesired inconsistencies by manually vetting their compiled documents, as it is difficult to determine how their documents will change otherwise.

\paragraph{Handling warnings and errors} 

Several discrepancies across \texlive versions were due to different ways of handling warnings and errors.
For example, in the 2021 distribution, \texttt{latexmk} would continue to attempt compilation on documents with syntax errors in their bibliography file; whereas in the 2022 distribution, \texttt{latexmk} would abort compilation.
While such differences might fall under undefined behaviour as they result from authors' syntax errors, acting on any warnings or errors output during compilation can help authors to avoid unexpected differences across \texlive versions.
Other discrepancies were directly indicated by compilation warnings: for instance, the font style inconsistencies in Figure~\ref{fig:sampleDiff:missingStyles01} corresponded to warnings stating that the missing fonts would be substituted.
Such warnings can help avoid authors avoid inconsistencies across different \tex engines.

\paragraph{Finding bugs}

Our approach identified various bugs in the \tex ecosystem through inter-version comparisons, as well as significant inconsistencies in cross-engine comparisons.
This pipeline may thus function as an automated differential testing pipeline for developing \tex-related software by providing a varied test corpus,
allowing developers to test how their software may interact with other software in the \tex ecosystem.
Due to the complexity of the \tex ecosystem, where authors tend to use a large variety of packages and document classes which may be updated independently of each other, 
such an approach would be highly beneficial in uncovering edge cases as it uncovers issues that might not be found in isolated tests.
This helps developers test how their software might be affected by, or result in, changes in the \tex ecosystem over a large number of diverse use cases.
Besides \tex-related differential testing campaigns, our analysis of the efficacy of various PDF comparison methods could inform the design of other testing approaches that similarly focus on human-readable documents.

%% file: sections/related-work.tex
\section{Related Work}

To our knowledge, there is no existing work on testing \tex engines or comparing \texlive distributions. 
In this section, we summarise related works that discuss similar approaches or adjacent goals.

\paragraph{Studying the \tex ecosystem} 

Much of \tex-related published work focuses on improving author experience through \latex packages~\cite{bless2023scikgtex, marris2024visualizing, Ohl_1995, Garc_a_Risue_o_2016}, automatically generating \latex code~\cite{vadimovich2013software, Kayal_2022, singh2018teaching, rosa2023recognizing, wang2019translating}, or developing \tex-adjacent software like syntax fixers~\cite{zhu2022eqfix} and command-line tools~\cite{Bar_2021,bizopoulos2023makefile}.
We found relatively less literature studying the \tex ecosystem itself.
The closest related work is a study on vulnerabilities in the \tex ecosystem~\cite{lacombe2021accept}, and a study on the role of \latex, Bib\tex, and metadata in publishing articles~\cite{bos2023latex}.

\paragraph{Comparing PDF documents} 

Comparing, and verifying the correctness of, PDF documents was a key challenge in this study.
An existing attempt at an automated PDF format checker was effective for all PDFs except those generated by \latex, as the underlying PDF parser could not retrieve sufficient font information in \latex-generated documents~\cite{YangTh:PDFFormatChecker}.
Another study compared PDF documents using image similarity algorithms to effectively detect inconsistencies~\cite{Kuchta2018:PDFCorrectness}.
However, this approach was not effective in our study as we had a higher threshold for permissible variation between documents, resulting in excessive false positives.

\paragraph{Identifying visual discrepancies}

One difficulty in this study was identifying output differences that were visible and significant.
\citet{Donaldson:TestingGraphicsShaderCompilers} used the chi-squared distance algorithm to compare image histograms and identify meaningful differences when testing graphics shader compilers, as slightly different results are allowed due to compiler optimisations.
We found similar methods unsuitable as differences in line breaks and hyphenation resulted in noise that obscured other types of inconsistencies.
Other methods of quantifying image similarity include summing the absolute distance between components of vectors~\cite{ImageSimilarityMetrics}, and pixel-wise algorithms like Structural Similarity Index (SSI) and Mean Square Error (MSE)~\cite{CompareMSESSI}.
These algorithms were unsuitable in our context as the different \tex engines often generated pixel-wise differences that were visually unnoticeable, resulting in false positives.

\paragraph{Fuzzing}

Fuzzing, short for ``fuzz testing'', is a test-case generation technique used in software testing~\cite{FuzzingStateOfTheArt}.
By creating random test cases as input to a program, fuzzing aims to uncover vulnerabilities or bugs through these malformed or unexpected inputs.
Fuzzing has been considered an effective method for testing software, and has been used to find bugs in UNIX utilities~\cite{FuzzingUNIX}, C compilers~\cite{Yang:FindingBugsCCompilers,McKeeman:DiffTesting}, MacOS applications~\cite{FuzzingMacOS}, file systems~\cite{FuzzingFileSystems} and more.
Improvements to fuzzing include using coverage-guided techniques to generate test cases that have greater code coverage, known as \textit{greybox fuzzing}~\cite{Fuzzing:CoverageGuided};
generating test cases that exercise different execution paths of the source program, known as \textit{whitebox fuzzing}~\cite{Fuzzing:Whitebox};
or using grammar-aware techniques to generate syntactically correct test cases, known as \textit{grammar-based fuzzing}~\cite{Fuzzing:GrammarBased}.
We briefly explored using generative and mutation-based fuzzing to generate test inputs, but ultimately found that arXiv provided suitably diverse usages of \tex.
Given the high availability of real-world \latex documents, we chose to use them as test inputs over fuzzing. 
Future work could further explore applying fuzzers to find crash bugs or to generate more obscure edge cases.

\paragraph{Differential testing}

Differential testing is a form of random testing that identifies a test case as a candidate to cause a bug if it results in different outputs in two (or more) comparable systems~\cite{McKeeman:DiffTesting}.
It has been used to find bugs in various domains, such as document processors~\cite{li2024finding}, software libraries~\cite{chen2015differential,godefroid2020differential}, program analysers~\cite{klinger2018differentially, Singhal_2022}, database systems~\cite{zheng2022finding, lin2022gdsmith, sotiropoulos2021data}, runtime engines~\cite{chen2016coverage, park2021jest, jiang2023revealing}, and most notably compilers~\cite{GrayC, lidbury2015many, sun2016finding, Yang:FindingBugsCCompilers}.
Given the similarity in methodology, our study can be seen as a form of differential testing.
However, different from these works, our outputs are neither well-defined nor not expected to be perfectly identical.
It was thus challenging to identify whether a discrepancy was expected or not.

\paragraph{Sanitising} 

Sanitisers are error-detection frameworks that check for various potential bugs, such as integer overflows~\cite{sanitiser:integeroverflow} or out-of-bound memory accesses~\cite{sanitiser180957, sanitiser7054186}.
Studies on differential testing on compilers have attempted to detect and reduce undefined behaviour in test cases using compiler sanitisers. 
For instance, the GrayC tool generates C programs and filters out invalid programs using Frama-C and Clang/LLVM compiler sanitisers~\cite{GrayC}.
While sanitising may have reduced the amount of false positives in our paper, we found sanitising to be impractical as we could not account for undefined behaviour in the individual \latex packages used by each test case.

%% file: sections/conclusion.tex
\section{Conclusion}

We have developed an automated pipeline for identifying inconsistencies between \tex engines and \texlive distributions.
Using this approach, we found a significant extent of inconsistencies throughout the \tex ecosystem, ranging from changes in default compilation flags and formatting in document classes to behaviours of \latex packages and the core \latex package itself.
We found two new bugs in \latex packages, along with five existing bugs and other obscured differences.
We characterised the inconsistencies found to demonstrate the typical ways in which \tex engines may differ beyond their explicitly listed differences, as well as what authors can expect from different versions of \texlive.
By focusing on the \tex ecosystem as a whole, we offer actionable and practical insights for authors of \tex documents to manage the constantly-evolving software and avoid undesirable outcomes. 
We believe this study will also be beneficial for developers by empirically demonstrating the relationship between different parts of the ecosystem, how inconsistencies manifest, and how it may affect their users.
Finally, our automated pipeline can be additionally used as a differential testing approach to gather a diverse test corpus and uncover more bugs and edge cases in the \tex ecosystem.

% XXX another key insight/contribution is looking at the system as a whole

%% file: sections/data-availability.tex
\section{Data Availability}

Our automated tool is available on GitHub at 
\url{https://github.com/jovyntls/inconsistencies-in-tex}.
This software, along with our experimental results and configurations for reproducibility, is archived for long-term storage at 
\url{https://zenodo.org/records/12669708}~\cite{tan_2024_12669708}.